\documentclass[floats,prd,aps,twocolumn,amssymb,nofootinbib,letterpaper, superscriptaddress,showkeys]{revtex4-1}
\usepackage{amsmath}
\usepackage{amsfonts}
\usepackage[utf8]{inputenc}
\pdfoutput=1
\usepackage[dvips]{graphicx}
\usepackage{graphicx}
\usepackage{color}
\usepackage{braket}
\usepackage{dcolumn}
\usepackage{bm,url}
\usepackage[linktocpage]{hyperref}
\usepackage{subfigure}
\usepackage{amsfonts}
\usepackage{comment}
\DeclareMathOperator\arctanh{arctanh}
\usepackage[usenames,dvipsnames,svgnames]{xcolor}  
\usepackage{hyperref}   
\definecolor{oxfordblue}{rgb}{0.0, 0.13, 0.28}
\definecolor{burgundy}{rgb}{0.5, 0.0, 0.13}
\definecolor{darkolivegreen}{rgb}{0.33, 0.42, 0.18}
\definecolor{darkblue}{rgb}{0,0,0.5}
\definecolor{richcarmine}{rgb}{0.84, 0.0, 0.25}
\definecolor{darkblue}{rgb}{0,0,0.5}
\definecolor{bluer}{rgb}{0.00,0.50,0.75}{}
\hypersetup{colorlinks=true, citecolor=blue, linkcolor=richcarmine,
urlcolor =magenta, filecolor=black}
\begin{document}
\newcommand{\tcr}{\textcolor{red}}
\newcommand{\tcb}{\textcolor{blue}}
\newcommand{\tcc}{\textcolor{cyan}}

\newcommand\be{\begin{equation}}
\newcommand\ee{\end{equation}}
\newcommand\bea{\begin{eqnarray}}
\newcommand\eea{\end{eqnarray}}
\newcommand\bseq{\begin{subequations}} 
\newcommand\eseq{\end{subequations}}
\newcommand\bcas{\begin{cases}}
\newcommand\ecas{\end{cases}}
\newcommand{\p}{\partial}
\newcommand{\f}{\frac}

\title{Anisotropic Generalized Polytropic Spheres: Regular 3D Black Holes}

\author{\textbf{Seyed Naseh Sajadi}}
\email{naseh.sajadi@gmail.com}
\affiliation{School of Physics, Institute for Research in Fundamental Sciences (IPM),\\ P. O. Box 19395-5531, Tehran, Iran}

\author{\textbf{Mohsen Khodadi}}
\email{khodadi@kntu.ac.ir}
\affiliation{Department of Physics, K. N. Toosi University of Technology,	P. O. Box 15875-4416, Tehran, Iran}

\author{\textbf{Orlando~Luongo}}
\email{orlando.luongo@unicam.it}
\affiliation{Universit\`a di Camerino, Via Madonna delle Carceri 9, 62032 Camerino, Italy.}
\affiliation{SUNY Polytechnic Institute, 13502 Utica, New York, USA.}
\affiliation{INAF - Osservatorio Astronomico di Brera, Milano, Italy.}
\affiliation{Istituto Nazionale di Fisica Nucleare, Sezione di Perugia, 06123, Perugia,  Italy.}
\affiliation{Al-Farabi Kazakh National University, Al-Farabi av. 71, 050040 Almaty, Kazakhstan.}

\author{\textbf{Hernando Quevedo}}
\email{quevedo@nucleares.unam.mx}
\affiliation{Instituto de Ciencias Nucleares, Universidad Nacional Aut\'onoma de M\'exico, Mexico.}
\affiliation{Dipartimento di Fisica and Icra, Universit\`a di Roma “La Sapienza”, Roma, Italy.}
\affiliation{Al-Farabi Kazakh National University, Al-Farabi av. 71, 050040 Almaty, Kazakhstan.}

\begin{abstract}
We model gravitating relativistic 3D spheres composed of an anisotropic fluid in which the radial and transverse components of the pressure correspond to the vacuum energy and a generalized polytropic equation-of-state, respectively. By using the generalized Tolman-Oppenheimer-Volkoff (TOV)  equation, and solving the complete system of equations for these anisotropic generalized polytropic spheres, for a given range of model parameters, we find three novel classes of asymptotically AdS black hole solutions with regular core. We show the regularity of the solutions using curvature scalars and the formalism of geodesic completeness.
Then, using the eigenvalues of the Riemann curvature tensor, we consider the effects of repulsive gravity in the three static 3D regular black holes, concluding  that their regular behavior can be explained as due to the presence of repulsive gravity near the center of the objects. Finally, we study the stability of the regular black holes under the flow of the energy through the Cauchy horizon.
\end{abstract}


\maketitle

\section{Introduction}
Since the 80s with the advent of the leading works by Gott and Alpert \cite{Gott:1982qg}, Giddings, Abbot and Kucha \cite{Giddings:1983es}, Deser, Jackiw, and t'Hooft \cite{Deser:1983tn, Deser:1983nh}, Brown, and Henneaux \cite{Brown:1986nw} and Witten \cite{Witten:1988hc}, gravity in lower-dimensions (specifically 3D i.e., $(2+1)$) became an active research topic for the theoretical community. Two reasons can be the motivation for this. First, taking care of some of the technical issues present in a wide range of problems in standard 4D gravity becomes notably easier in lower dimensions, (e.g., see seminal papers \cite{Mann:1991qp, Mann:1995eu, Elizalde:1992zm}). From the point of quantum gravity, one reason that handling General Relativity in lower dimensions is significantly simpler is that it, in essence, becomes a topological theory without any propagating local degrees of freedom \cite{book}. This is different from 4D since the Weyl tensor is zero and subsequently, the gravitational field spacetime has no dynamic degrees of freedom, resulting in curvature being   produced only by matter\footnote{Another way to arrive at this absence of degrees of freedom is through the counting argument fo the  so-called canonical geometrodynamics, which can be found in \cite{Giddings:1983es} (see also \cite{book}).}. 
Second, there are some physical systems such as cosmic strings and domain walls that effectively recommend motion on lower--dimensional geometries \cite{Book}. 3D gravity has also this advantage that let us evaluate the connection between gravity and  gauge field theories, since it can be formulated in the form of a Chern-Simons theory \cite{Achucarro:1986uwr}. Even though General Relativity in 3D can serve as an effective laboratory for investigating conceptual issues, it is not free of some issues. In this direction, one can mention the lack of the Newtonian limit, meaning that there is no gravitational force between
masses \cite{Barrow}. Although adding topological and higher-derivative terms to Einstein's gravity leads to the propagation of local degrees of freedom, they cause the appearance of some other problems in the holographic context \cite{Deser:1981wh, Bergshoeff:2009hq, Bergshoeff:2014pca, Setare:2014zea, Ozkan:2018cxj}.

3D Einstein's gravity suffers from a lack of variety of solutions since gravity is trivial \cite{Ida:2000jh}. In 3D flat spacetime, the only solution with a horizon is the flat space cosmology \cite{Bagchi:2013lma, Setare:2020mej}. The other solution of this kind is the kink-like solution with gravitational field, which addresses a conical space with a deficit angle describing the particle's mass. Despite the triviality of gravity in 3D, Ba\~nados, Teitelboim, and Zanelli (BTZ) have shown that in the locally anti-de Sitter (AdS) spacetime, there are black hole solutions in addition to the particle-like solution \cite{Banados:1992wn}. 
The negative sign of the cosmological constant is important because in the case of $dS_3$ no black holes can exist \cite{emparan2022black}. 
Although cosmological observations in 4D are compatible only with a dS spacetime, the  study of black hole configurations in 4D and 3D implies the consideration of AdS spacetimes 
for pure theoretical reasons. 

\color{black}
The BTZ black hole spacetime, in essence, is obtained by identifying certain points of the AdS spacetime. As for the significance of 3D Einstein's gravity to study black holes, it is essential to mention that it is indeed the lowest dimension allowed.
Rotating BTZ\footnote{One can find the electric and magnetic versions of the rotating BTZ black hole in Refs. \cite{Martinez:1999qi}, and \cite{Dias:2002ps}, respectively.} is characterized by the mass, angular momentum, and cosmological constant, providing some joint characteristics with Kerr's black hole in conventional four-dimensional Einstein gravity. Actually, it can be considered as an endpoint of the gravitational collapse in AdS$_3$ spacetime. It is interesting to note that as to the origin of entropy in BTZ black holes, clues may be found in the formulation of 3D gravity based on the Chern-Simons theory \cite{Carlip:1994gy}. 
Concerning the status of singularity in BTZ black holes, it needs to be mentioned that the rotating BTZ, regarding existence of closed time-like curves, is free of curvature singularities at the origin, while its non-rotating version is singular \cite{Banados:1992gq}. Similar to well-known conical singularities discussed in \cite{Deser:1983tn, Gott:1982qg}, in the absence of angular momentum, the BTZ black hole is an example of a singular spacetime \footnote{In Ref. \cite{Pantoja:2002nw}, it has been demonstrated that the non-rotating BTZ black hole metric, in essence, belongs to those classes of metrics that are semi-regular.}. The noteworthy point is that the conical singularity differs from the canonical singularity in that it does not cause a physical divergence of the curvature of spacetime.
That is, there is no danger to spacetime and to the validity of causality from the side of the conical singularity \cite{Chen:2019nhv}.

Motivated by the BTZ black hole, the extended 3D black hole solutions and some of their applications have received much attention from theoretical physics researchers in recent years, e.g., see \cite{Carlip:1995qv, Rincon:2018sgd, Panotopoulos:2018pvu, Podolsky:2018zha, Sharif:2019mzv, Ali:2022zox, Priyadarshinee:2023cmi, Karakasis:2023ljt}. In this direction, one can find some studies which address $(2+1)$-dimensional solutions of gravitational field equations beyond General Relativity, including electrically charged black holes, dilatonic black holes, black holes arising in string theory, black holes in topologically massive gravity and warped-AdS black holes \cite{mann1994, Parsons:2009si, Hennigar:2020fkv, Hennigar:2020drx, Sheykhi:2020dkm, Eiroa:2020dip, Karakasis:2021lnq, Nashed:2021ldz, Karakasis:2021ttn}. Therefore, it is no exaggeration to say that the idea of gravity in 3D has become much more exciting after the introduction of the BTZ black hole. On the other hand, the absence of gravitational waves in BTZ gives us the chance to have a solvable model that includes quantum properties of black holes since its existence in 4D, due to nonlinear interactions, does not let us an exact solution of any system that includes quantum gravity \cite{Witten:2007kt}.

By involving BTZ black holes in various scenarios and exposing them to some interesting phenomena in recent years (for instance, see Refs. \cite{Clement:1995zt, Cadoni:2008mw, Quevedo:2008ry, Liu:2011fy, Hendi:2012zz, Sheykhi:2014jia, Ferreira:2017cta, Dappiaggi:2017pbe, Fujita:2022zvz,deOliveira:2022csc, Poojary:2022meo, Furtado:2022tnb, Yu:2022bed, Basile:2023ycy, Baake:2023gxx, Konewko:2023gbu, Du:2023nkr, Zhou:2023nza, Chen:2023xsz, Devi:2023fqa, Jeong:2023zkf, Fontana:2023dix, Birmingham:2001pj, Ren:2010ha, Momeni:2015iea, Eberhardt:2019ywk, Eberhardt:2020akk, Balthazar:2021xeh}), more details of its physical function have been now revealed to us. In the meantime, what attracts our attention is the attempt to provide 3D black hole solutions that are free of the central ($r=0$) singularity, see Refs. \cite{He:2017ujy,Cataldo:2000ns, Myung:2009atx,Bueno:2021krl, Jusufi:2022nru, Jusufi:2023fpo,Karakasis:2023hni}. The fact that the resolution of the singularity problem is not restricted to 4D, but its existence in any number of dimensions of spacetime signals a pathological behavior, is the key motivation to extend regular black holes to 3D. In other words, to have a well-defined spacetime in any dimension, the gravitational singularity at the center of black holes must be controlled. In line with this contribution, we extend the above catalog (Refs. \cite{He:2017ujy, Cataldo:2000ns, Myung:2009atx, Bueno:2021krl, Jusufi:2022nru, Jusufi:2023fpo, Karakasis:2023hni}) with a new family of 3D black hole solutions, including a regular center that asymptotically recovers BTZ black holes. Although regular black holes are commonly the solutions of gravity coupled to nonlinear electrodynamics, here in the framework of Einstein's gravity \cite{Lan:2023cvz,Sebastiani:2022wbz}, by taking into account the generalized Tolman-Oppenheimer-Volkoff (TOV) equation for an anisotropic generalized polytropic fluid, we present different classes of new black hole solutions in AdS$_3$ spacetimes that are free of the central singularity.
More precisely, these solutions are obtained by including an anisotropic-barotropic fluid obeying a generalized polytropic equation-of-state (GPEoS) within Einstein's gravity. 
It is necessary to emphasize that we checked the regularity of the black hole solutions using both the finiteness of some curvature tensor-based scalars, such as the Ricci and Kretschmann scalars, and the geodesic completeness of the corresponding spacetimes.
	
We complete the analysis of the solutions derived in this work by studying  repulsive effects, shadows, and the stability of the Cauchy horizon.

\section{anisotropic TOV equation}\label{sec2}
Let us take the following static, circularly symmetric metric for describing a gravitating relativistic $(2+1)$-dimensional object in global coordinates ($x^{\mu}=(t,r,\phi)$)
\begin{equation}\label{metric}
ds^{2}=-f(r)dt^{2}+g(r)dr^{2}+h(r)d\phi^2,
\end{equation} 
where $f$, $g$, and $h$ denote the unknown functions with  
$t,r \in (-\infty, +\infty)$, and $\phi \in [0,2\pi]$.
The field equations are  
\begin{equation}
G_{\mu \nu}+\Lambda g_{\mu \nu}=8\pi G T_{\mu \nu},
\end{equation}
where $\Lambda=-1/\ell^2$ is a negative cosmological constant. By considering an anisotropic fluid, the 3D energy-momentum tensor reads off as
\begin{equation}
T^{\mu}{}_{\nu}=\mbox{diag}(-\rho,P_{r},P_{t}),
\end{equation} where $\rho$, $P_{r}$, and $P_{t}$ are the energy density and the radial and transverse components of the pressure, respectively. We obtain the generalized TOV equation from 
$\nabla_{\mu}T_{r}^{\mu}=0$, which leads to
\begin{equation}\label{eqTOV1}
	\dfrac{dP_{r}}{dr}=-\dfrac{1}{2}
	\left[\dfrac{(\rho+P_{r})f^{\prime}}{f}-\dfrac{
		\Delta h^{\prime}}{h}\right],
\end{equation} with the anisotropic function $\Delta=P_t-P_r$. The components of the field equations take the following form
\begin{align}
tt:&\;\;\;\;\dfrac{h^{\prime 2}}{h^{2}}-\dfrac{2h^{\prime\prime}}{h}+\dfrac{g^{\prime}h^{\prime}}{hg}+\dfrac{4g}{\ell^2}-4g\rho= 0,\label{a}\\
rr:&\;\;\;\;\;\dfrac{f^{\prime}h^{\prime}}{fh}-\dfrac{4g}{l^2}-4P_{r}g=0,\label{b}\\
\phi\phi :&\;\;\;\;\dfrac{f^{\prime 2}}{f^{2}}-\dfrac{2f^{\prime\prime}}{f}+\dfrac{g^{\prime}f^{\prime}}{fg}+\dfrac{4g}{\ell^2}+4gP_{t}=0.\label{c}
\end{align}
The above system of equations comprises six independent
variables $(\rho,P_r,P_t,f(r),g(r),h(r))$ with three equations. We now proceed to handle the above system of equations. Inspired by Refs. \cite{Chavanis:2012pd, Chavanis:2012kla,Kiselev:2002dx,Capozziello:2022ygp,Giambo:2023zmy,Debnath:2004cd}, in which to present various cosmological models of the early (high-density) and late (low-density) universe, the author utilized GPEoS for the radial pressure $P_r(\rho)$, we here extend it  to the transverse component too \cite{Sajadi:2016hko, Riazi:2015hga} 
\begin{align}\label{eqq9}
P_{r}=w \rho+w^{\prime}\dfrac{\rho^{n}}{\rho_{0}^{n-1}},\;\;\;\;\;\;
P_{t}=w_{1} \rho+w_{2}\dfrac{\rho^{m}}{\rho_{0}^{m-1}},
\end{align}
where, $\rho_{0}$ is the central density, $w,w_{1}$ are the dimensionless EoS parameters and $w^{\prime}, w_ 2$ together with $n,m$ denote the polytropic constants and indices, respectively. We will find out that the key results released throughout this work, in essence, come from this extension, i.e., $P_t(\rho)$. Providing some notes on the existing EoS can be helpful. 

Clearly, it is a combination of the linear EoS and polytropic EoS, which connects the pressure and density via a power-law. The latter is usually expected to be relevant in the so-called polytropes, i.e., self-gravitating gaseous spheres that were, and still are, very useful as crude approximations to more realistic stellar models \cite{chander}.
	They are usually utilized in various astronomical situations to model compact objects in two very different types of regimes, i.e., Newtonian (non-relativistic) and General Relativity \cite{Herrera:2013fja}. Concerning the latter, i.e., enough compact objects, the polytropic EoS has been widely used, \cite{Nilsson:2000zg, Maeda:2002br, Sa:1999yf, Lai:2008cw, Setare:2015xaa, Contreras:2018dhs, Abellan:2020nkl} (see also references therein).
	Although the EoS of very compact objects such as neutron stars, is still unknown, in recent years a lot of insight has been obtained by using the polytropic EoS to integrate the stellar equations of structure. So, this justifies the use of GPEoS, such as  (\ref{eqq10}), in black hole configurations. 
 Presumably, a black hole might have a star-like internal structure whose density and pressure, can be varied from the center outwards (not necessarily in a linear manner). In other words, it is expected to be suitable to use a polytropic EoS whose behavior changes from soft to complex at low and high densities, respectively. Recently, by investigating the formation of regular Hayward and Bardeen black holes arising from the gravitational collapse of a massive star, it was proposed in \cite{Shojai:2022pdq} that the EoS of a regular black hole may be polytropic.

	Besides, unlike the usual assumption that the pressure of the fluid under consideration is isotropic, i.e., $P_r=P_t$, here we consider the case of anisotropy, i.e., $P_r\neq P_t$. The existence of anisotropy in the fluid causes additional pressure against gravity, allowing the compact object to be more stable. Again according to the presumed similarity of the internal structure of a black hole and star, one can find some indications that allow us to take anisotropic fluid into account.

 First, local pressure anisotropy may result from a wide range of physical phenomena of the type we would expect to find in compact objects \cite{Mimoso:2013iga, Nguyen:2013nka}. Second, due to the presence of viscosity in highly dense mediums, anisotropy fluids are expected  to be inevitable \cite{Drago:2003wg, Herrera:2004xc}. Third, in cosmology, anisotropy in the fluid pressure  has different functions, such as generating natural inhomogeneities at small scales and explaining the matter density power spectrum at short wavelengths \cite{Cadoni:2020izk}. In Ref. \cite{Freitas:2013nxa}, the GPEoS  has been utilized to evaluate primordial quantum fluctuations to provide a universe with a constant energy density at the origin.

In the following, to obtain solutions with $T^{t}_{t}=T^{r}_{r}$ resulting in $f=g^{-1}$, we have to set $\omega=-1,\omega^{\prime}=0$ for the dimensionless EoS parameters of the radial pressure in (\ref{eqq10}).  As a result, the  anisotropic fluid consists, in essence,  of the following radial and transverse components of the pressure 
    \begin{align}
\label{eqq10}
&P_{r}=- \rho,\;\;\;\;\;\;\\ 
&P_{t}=w_{1} \rho+w_{2}\dfrac{\rho^{m}}{\rho_{0}^{m-1}},\label{eqq11}
    \end{align} 
which address the vacuum energy and GPEoS, respectively. 

The radial pressure is associated with a de Sitter phase, as consequence of the field equations. This contribution prevails not only in the de Sitter core but also inside the entire black hole.

Despite the absence of the polytropic constant $w'$ throughout our analysis, it would be interesting to mention that its origin may come from Bose-Einstein condensates with repulsive (attractive) self-interaction if $w'>0 (<0)$ \cite{Chavanis:2012pd, Chavanis:2012kla}.
Now, the field equations, Eqs. (\ref{a})-(\ref{c}), re-express as follows
\begin{align}
	tt:&\;\;\;\;\dfrac{fh^{\prime 2}}{h^{2}}-\dfrac{2fh^{\prime\prime}}{h}-\dfrac{f^{\prime}h^{\prime}}{h}+\dfrac{4}{\ell^2}-{4\rho}= 0,\\
	rr:&\;\;\;\;\dfrac{f^{\prime}h^{\prime}}{h}-\dfrac{4}{\ell^2}+4\rho =0,\\
	\phi\phi :&\;\;\;\;-f^{\prime\prime}+\dfrac{2}{\ell^2}+2P_{t}=0.\label{eqq12}
\end{align}
By adding the $tt$ and $rr$ components of the field equations, one can obtain $h(r)$ as follows
\begin{equation}
	h(r)=\dfrac{1}{4}h_{0}r^{2}+\dfrac{1}{2}h_{0}h_{1}r+\dfrac{1}{4}h_{1}^{2} .
\end{equation}
By setting $h_{1}$ equal to zero and redefining $\phi$ and replacing it in \eqref{eqTOV1}, we get
\begin{equation}\label{eqq8}
	r\rho^{\prime}+\rho+P_{t}=0.
\end{equation}
By putting $P_t$ from \eqref{eqq11} into \eqref{eqq8}, one can obtain $\rho$ as follows
\begin{equation}\label{eqqrho}
	\rho(r)=\dfrac{\rho_{0}}{\left[c_{1}\left(\rho_{0}r^{\omega_{1}+1}\right)^{m-1}-\dfrac{\omega_{2}}{\omega_{1}+1}\right]^{\frac{1}{m-1}}},
\end{equation}
where $c_{1}$ is an integration constant. Near the origin, the energy density is maximal and reads 
\begin{equation}\label{eqqnear}
	\rho(r\to 0)\approx \dfrac{\rho_{0}}{\left(\dfrac{-\omega_{2}}{w_{1}+1}\right)^{\frac{1}{m-1}}},
\end{equation}
{which implies a dS vacuum core.} For large values of $r$, Eq. \eqref{eqqrho} takes the following form 
\begin{equation}\label{eqqfar}
\rho(r\to \infty) \approx \dfrac{1}{c_{1}r^{w_{1}+1}},
\end{equation} indicating that the integration constant $c_{1}$ is related to the linear part of $P_t(\rho)$. As expected, $\omega_{2}$ appears only in the energy density of near the center, i.e., Eq. (\ref{eqqnear}).

An interesting feature emerging from the solutions in Eqs.  \eqref{eqqnear} and \eqref{eqqfar} is that in the former appears the imprint of both linear and power-law parts of $P_t$, as given in Eq. \eqref{eqq10}, while in the latter, just linear part. This is reasonably a consequence of Eq. \eqref{eqq10}. Indeed,  nonlinear effects are usually expected to be dominant in high-density mediums, while at low-density, the linear approximation works well. 

In general, it can be concluded that the integration constant $c_1$ actually depends on the linear part of $P_t(\rho)$ in Eq. \eqref{eqq11}, and not on the nonlinear part.
In the following, by setting some selected values for the free parameters $(m, \omega_{1}, \omega_{2})$, we will present some 3D black hole solutions which are free of singularities. To provide non-singular solutions, we will choose positive values for $m$ and $\omega_{1}$. This situation is reminiscent of the non-singular early universe model presented with GPEoS in Ref. \cite{Chavanis:2012pd}. Since we are interested in analytic expressions, we will present only a few  black hole solutions.

\begin{figure*}[ht!]
	\centering
	\includegraphics[width=0.95\columnwidth]{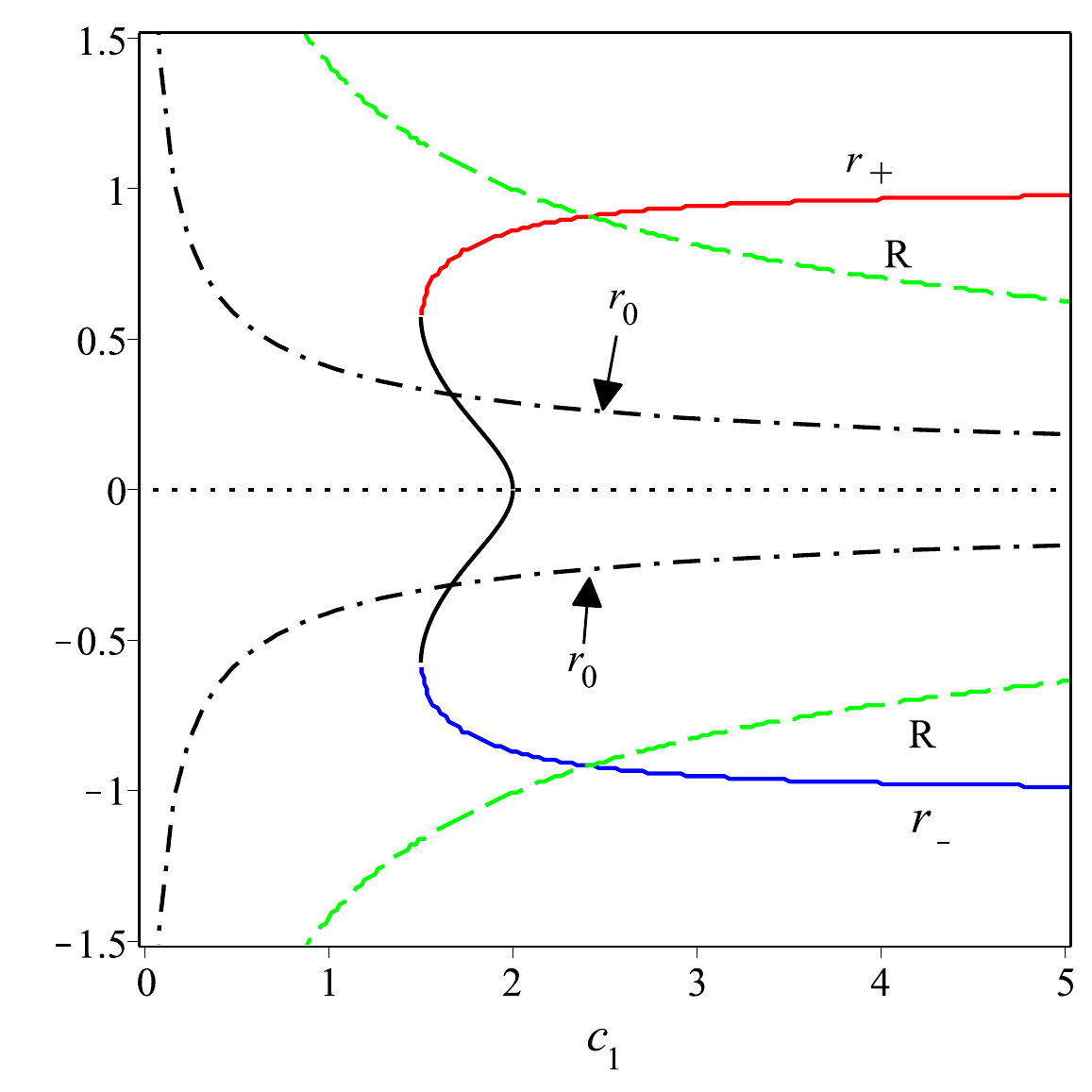}~~~~~~~~~~
	\includegraphics[width=0.95\columnwidth]{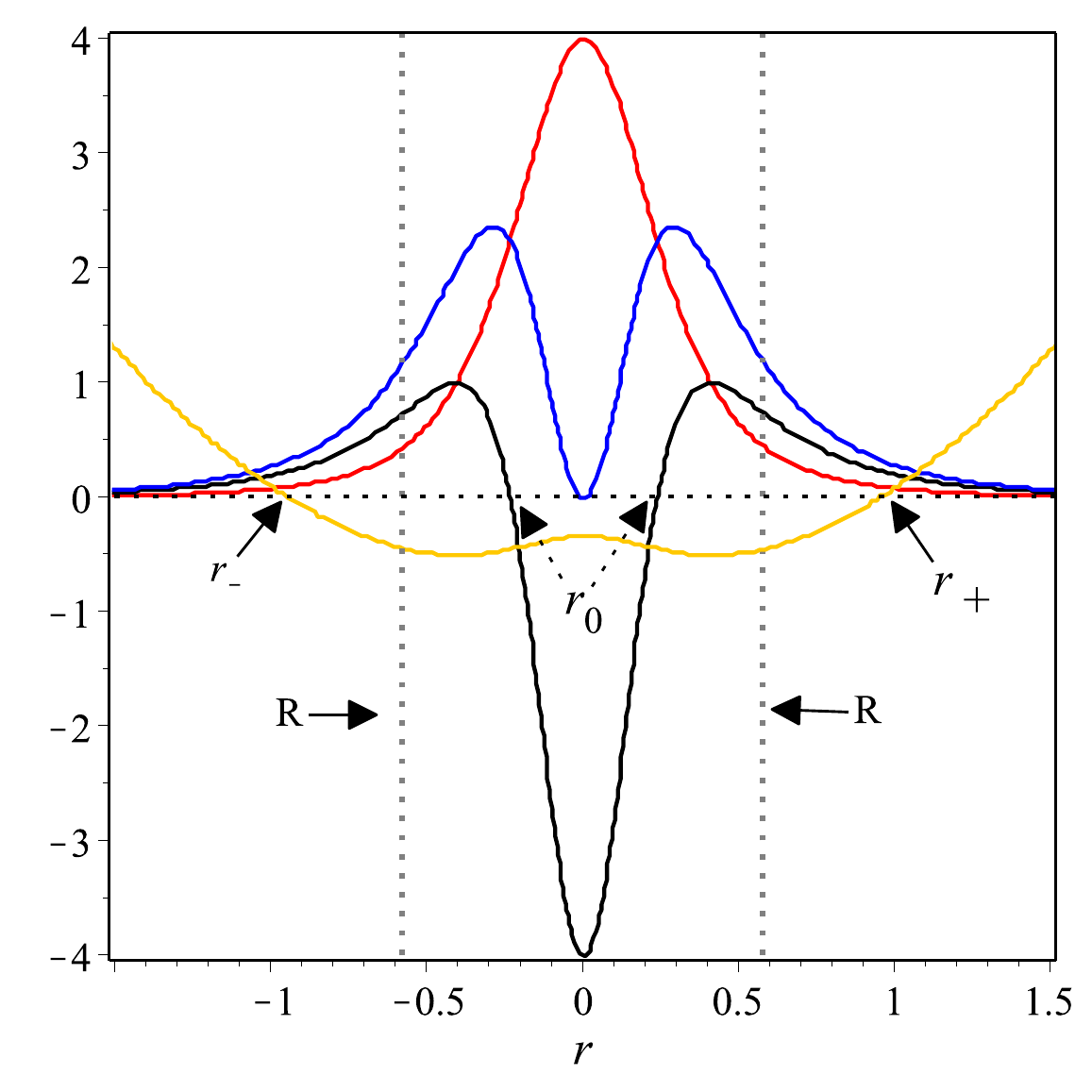}
	\caption{The behavior of $r_{\mp,0}$, and $R$ in terms of $c_{1}$ (left panel), and the behavior of $\rho$ (red), $\rho+P_{t}$ (blue), $\rho+P_{r}+P_{t}$ (black) and $f(r)$ (orange) in terms of $r$ (right panel) for the first black hole solution with the free parameters: $m=3/2, w_{1}=3,w_2=-2$, and  $\rho_{0}=\ell=1$. } 
	\label{rcplott}
\end{figure*}

\begin{figure}[ht!]
	\centering
 \includegraphics[width=0.95\columnwidth]{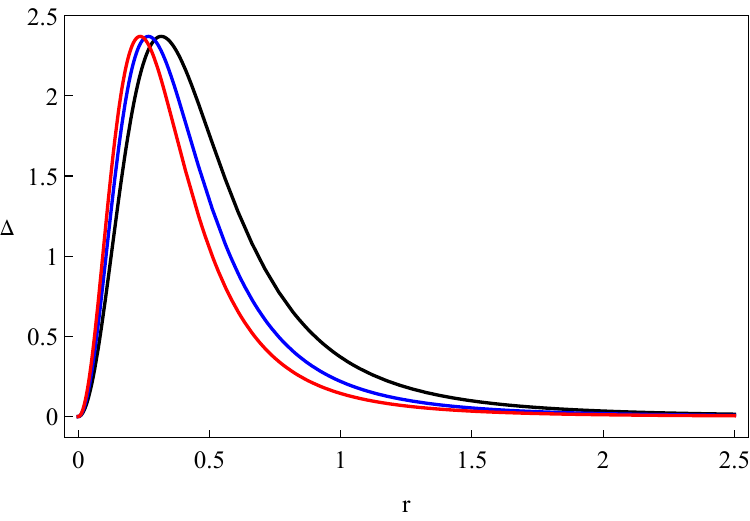}~~~
	\caption{The behavior of $\Delta-r$ for the first black hole solution with the same numerical values used in Fig. \ref{rcplott}. For the model parameter $c_1$, we have set numerical values $c_1=2.5,~3.5$, and $4.5$ in the black, blue, and red curves, respectively.} 
	\label{First}
\end{figure}

\subsection{The first solution: $m=3/2$, $w_{1}=3$}

To motivate the choices of constants, we notice that any value given to $m$ and $w_{1}$ leads to a possible solution. However, not all solutions yield a finite value for black hole mass. Hence, in our choices of constants, we present those values that provide a finite black hole mass.

\noindent Thus, inserting Eq. \eqref{eqqrho} into Eq. \eqref{eqq12}, one can obtain $f(r)$ as follows
\begin{equation}\label{first}
	f(r)=c_{2}+\dfrac{r^2}{\ell^2}+\dfrac{4\sqrt{\rho_{0}}}{c_{1}(4c_{1}\sqrt{\rho_{0}}r^2-w_{2})}.
\end{equation}
This solution describes the gravitational field of a black hole if there exists an outer horizon  located outside the radius of the compact object, i.e., $r_+ > R$.
The gravitational mass $M(r)$ inside a circle of radius $r$ is given by 
\begin{equation}
	M=2\pi\int_{0}^{\infty} r\rho(r)dr=-\dfrac{4\pi\sqrt{\rho_{0}}}{c_{1}\omega_{2}}.
\end{equation}
We see that the central density depends on the square of the asymptotic value of the mass. Moreover, with the choice of constants that we made, a finite value of mass, as requested above, is found.

It is clear that both linear and power-law parts of $P_t(\rho)$ in Eq. \eqref{eqq11} contribute to the gravitational mass so that it is positive ($M>0$) if $c_1\omega_2<0$ i.e.,  $\omega_2<0, c_1>0$, or $\omega_2>0, c_1<0$. If we assume that the above mass is equal to $M=\rho_{0}R^{2}$, then we obtain  
\begin{equation}
	R=\sqrt{-\dfrac{4\sqrt{\rho_{0}}}{c_{1}\omega_{2}\rho_{0}}},
\end{equation}
which can be considered as the effective radius of the object.
On the other hand, the roots of the metric function are given by
\begin{align}\label{eqf22}
	r_{\pm}=&\pm\dfrac{1}{4c_{1}}\sqrt{\dfrac{2c_{1}\omega_{2}-8c_{2}\ell^2c_{1}^{2}\sqrt{\rho_{0}}\mp 2\sqrt{Y}}{\sqrt{\rho_{0}}}}\\
	Y=&c_{1}^{2}\omega_{2}^{2}+8c_{1}^{3}c_{2}\omega_{2}\ell^{2}\sqrt{\rho_{0}}+
	16\rho_{0}\ell^{4}c_{1}^{4}c_{2}^{2}-64\rho_{0}\ell^{2}c_{1}^{2},\nonumber
\end{align} for $Y>0$, the $r_+$ address the event horizon, and $r_-$ is the Cauchy horizon. i.e, $0<r_{-}<r_{+}$. For $c_{1}\gg 1$, the horizon radii are given as
\begin{equation}
	r_{\pm}\sim \pm \ell\sqrt{-c_{2}}\mp\dfrac{\sqrt{-c_{2}}}{2\ell c_{1}^{2}c_{2}^{2}}+\mathcal{O}(c_{1}^{-3}),\;\;\;\;\;c_{1}\gg 1.
\end{equation}
The extremal roots in which the roots degenerate to a single one are obtained as ($Y=0$)
\begin{equation}
	c_{2}=-\dfrac{\omega_{2}+8\ell\sqrt{\rho_{0}}}{4c_{1}\ell^2\sqrt{\rho_{0}}}\;\to\;r_{ext}=0,\pm\dfrac{1}{2}\sqrt{\dfrac{w_{2}+4\ell\sqrt{\rho_{0}}}{c_{1}\sqrt{\rho_{0}}}}.
\end{equation}
In the further analysis, we will not consider the case $w_2>0, c_ 1<0$, since $r_{ext}$ becomes imaginary. Fig. \ref{rcplott} (left panel) clearly shows that $r_+>R$ for  $c_1$ bigger than a given value,  indicating that the solution in Eq. \eqref{first} describes a 3D black hole.

To find the physical significance of the parameters $c_1$ and $c_2$, we consider the behavior of the solution \eqref{first}. Hence, at $r\to\infty$ and $r\to 0$ we have 
\begin{equation}\label{1}
	f(r)\sim c_{2}+\dfrac{r^{2}}{\ell^2}+\dfrac{1}{c_{1}^{2}r^{2}}+\mathcal{O}(r^{-4}),
\end{equation}
and
\begin{equation}\label{2}
	f(r)\sim c_{2}-\dfrac{4\sqrt{\rho_{0}}}{c_{1}w_{2}}+\left(\dfrac{1}{\ell^2}-\dfrac{16\rho_{0}}{w_{2}^{2}}\right)r^{2}+\mathcal{O}(r^{4}),
\end{equation}
respectively. By comparing the Eq. \eqref{1} with a static BTZ black hole, it is not hard to figure out that 
$c_{1}$ corresponds to a  parameter with dimension $[c_{1}]=L^{-1}$ and $c_2$ is the dimensionless mass parameter $\mathcal{M}$.
By re-expressing Eq. \eqref{2} in the following form
\begin{equation}\label{22}
f(r)\sim c_{2}^{\prime}+\Lambda_{eff}r^2+\mathcal{O}(r^4)
\end{equation} one can interpret the coefficient of $r^{2}$ in the form of an effective cosmological constant
\begin{equation}
\Lambda_{eff}=\dfrac{16\rho_{0}}{\omega_{2}^{2}}-\dfrac{1}{\ell^2}.
\end{equation}
For $\omega_{2}^{2}/(16\rho_{0}\ell^2)<1$, then $\Lambda_{eff}>0$, and $c^{\prime}_{2}=1$ indicating that the solution at hand near the origin is the dS vacuum.
It is notable that the origin of the dS vacuum inside the black hole, in essence, comes from the power-law part of  $P_t(\rho)$ in GPEoS, Eq. \eqref{eqq11}. 

By calculating two curvature invariants, the Ricci and Kretschmann scalars, near the origin
\begin{align}
&	\lim_{r\to 0}\mathcal{R}=-\dfrac{6}{\ell^2}+\dfrac{96\rho_{0}}{w_{2}^{2}}=6\Lambda_{eff},\\
&\lim_{r\to 0}\mathcal{K}=\lim_{r\to 0}\mathcal{R}_{a b c d}\mathcal{R}^{a b c d}=\dfrac{12(-w_{2}^{2}+16\rho_{0}\ell^{2})^{2}}{\ell^{4}w_{2}^{4}}=12\Lambda_{eff}^{2},
\end{align}
one can see that they are finite at $r=0$, meaning that the first black hole solution is free of curvature singularities at the center.
From Fig. \ref{rcplott} (right panel), one can conclude that the weak energy condition (WEC) i.e., $\rho>0$ and $\rho+P_t>0$
is satisfied everywhere, while the strong energy condition (SEC), $\rho+P_r+P_t>0$, is satisfied everywhere except at
\begin{equation}\label{r0}
	r_{0}=\pm\dfrac{1}{6}\sqrt{\dfrac{-3w_{2}}{c_{1}\sqrt{\rho_{0}}}},
\end{equation}
which is located near the origin. 
From the right panel of Fig. \ref{rcplott}, it is also clearly evident that $r_+>R$. Equation \eqref{r0} carries the message that the power-law part of $P_t(\rho)$ in GPEoS caused the SEC physics to be forced into a small region in the deep core i.e., $r<r_{0}$. This is interesting in the sense that it also happens in some regular 4D black holes so that one can take $r_0$ as a characteristic length associated with the Planck scale (e.g., see \cite{Carballo-Rubio:2018pmi, Khodadi:2022dyi}). 

As a double-check to ensure the existence of a vacuum near $r=0$, one can evaluate the anisotropic function $\Delta$ for the first solution. Indeed the anisotropic function is a suitable term to illustrate the internal structure of the black hole (or any relativistic stellar objects). The behavior of $\Delta-r$ in Fig. \ref{First} clearly shows that although $\Delta$ reaches a maximum inside the black hole at a certain distance from the origin, it disappears as the center $r=0$ is approached, meaning that around the origin we are dealing with a vacuum.

\begin{figure*}[ht!]
\centering
	\includegraphics[width=0.95\columnwidth]{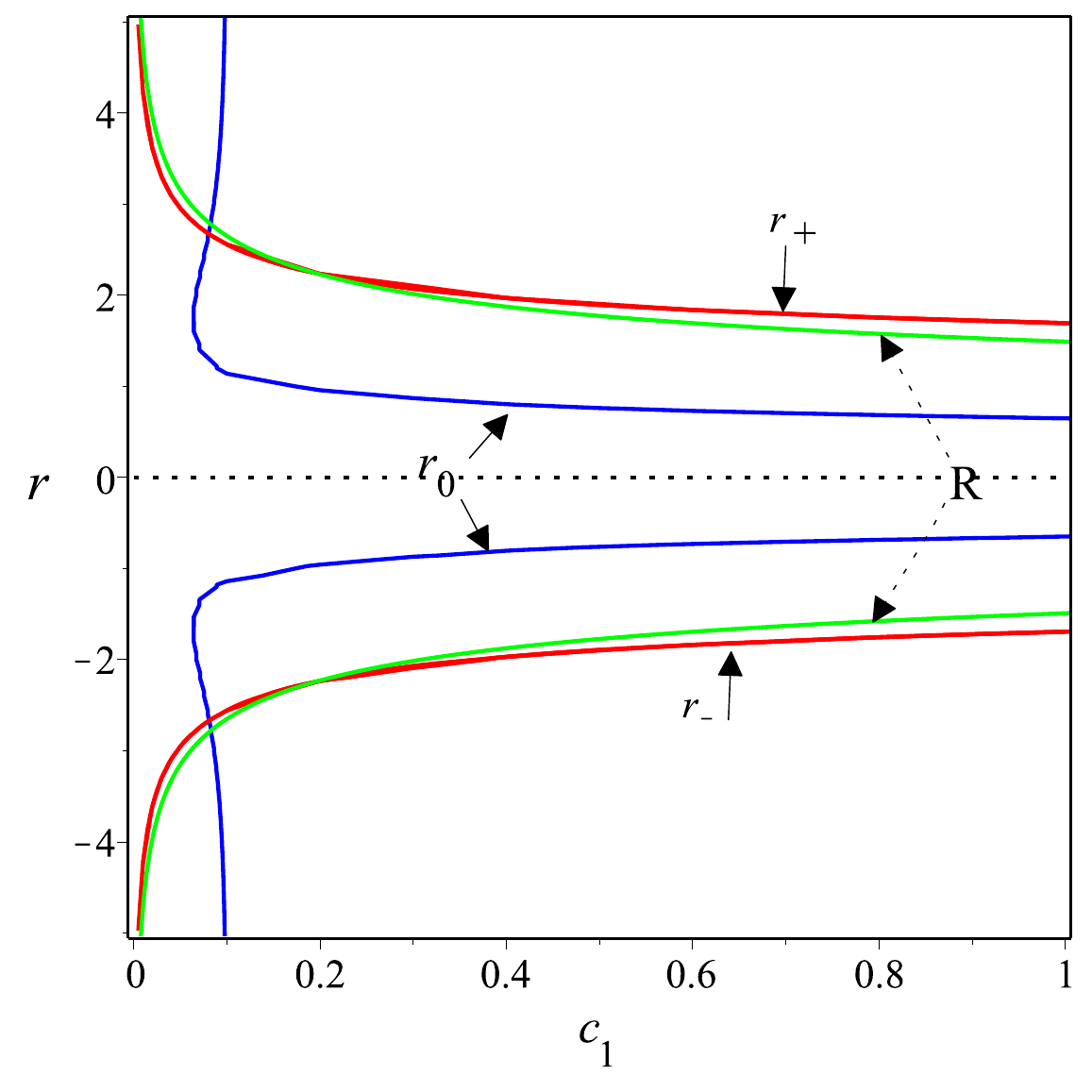}~~~~~~~~~~
	\includegraphics[width=0.95\columnwidth]{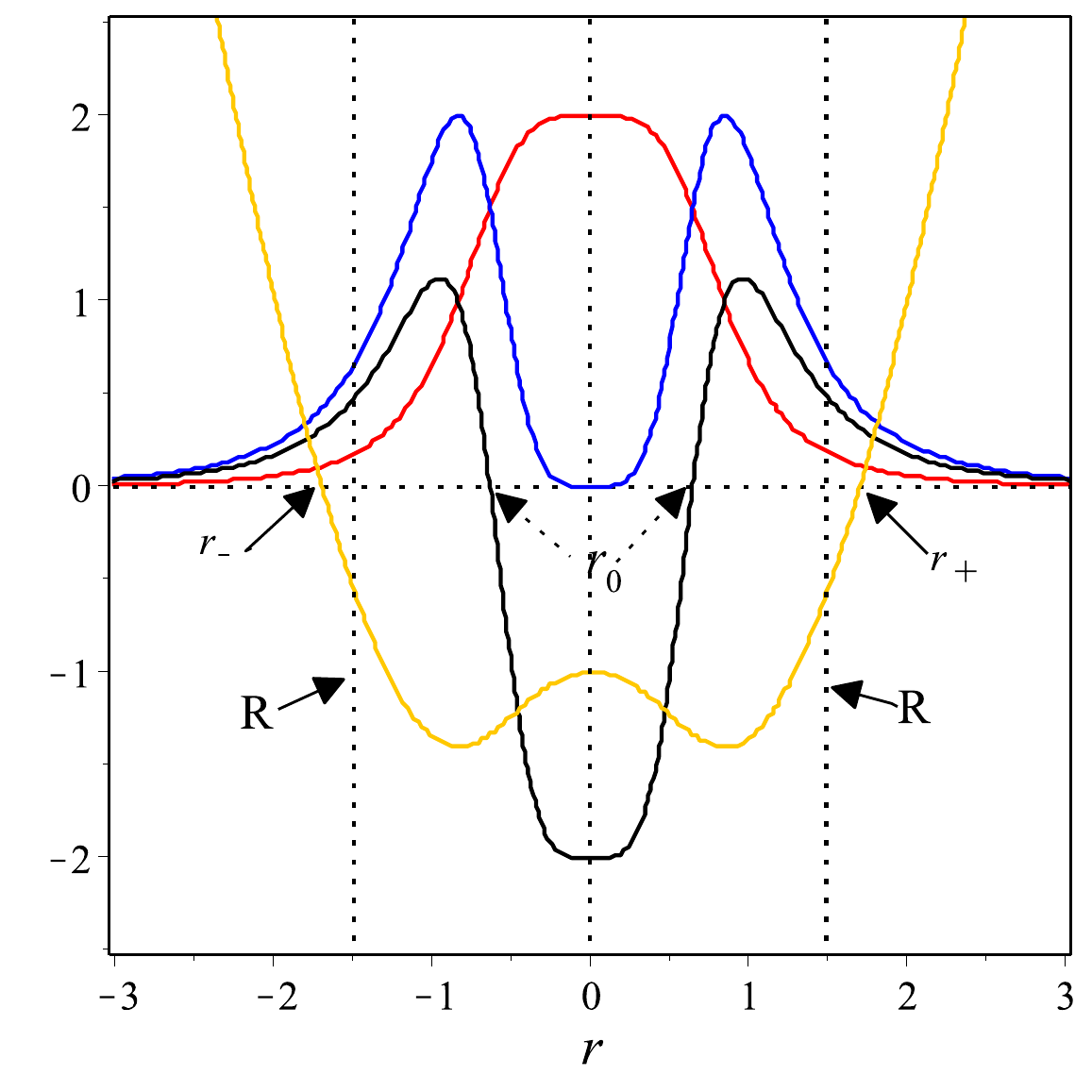}
	\caption{Same as Fig. \ref{rcplott} but for the second black hole solution with free parameters  $m=2, \omega_{1}=3, \omega_2=-2$, and $\rho_{0}=\ell=1$.
	} 
	\label{fplot1}
\end{figure*}

\begin{figure}[ht!]
	\centering
	\includegraphics[width=0.95\columnwidth]{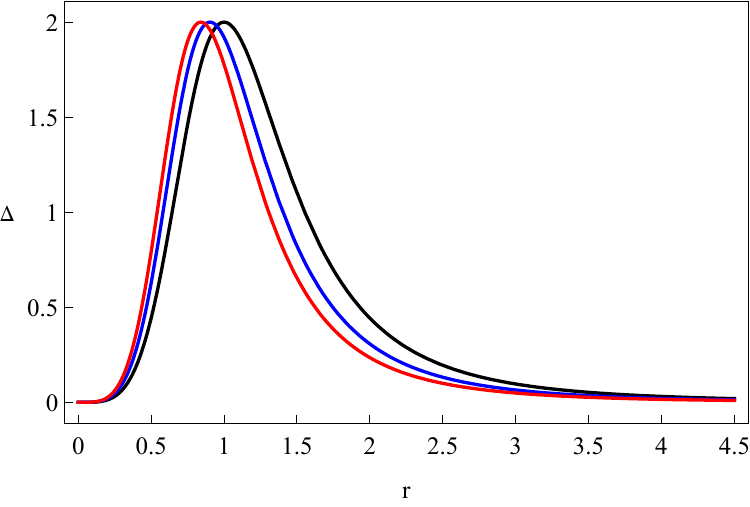}~~~
	\caption{The behavior of $\Delta-r$ for the second black hole solution with the same numerical values used in Fig. \ref{fplot1}. For the model parameter $c_1$, we have set numerical values $c_1=0.5,~0.75$, and $1$ in the black, blue, and red curves, respectively.} 
	\label{Second}
\end{figure}

{According to Appendix \ref{app1}, the computation of the curvature eigenvalues, $\lambda_{i}$, for this metric yields }
\begin{align}
    \lambda_{1}&=\dfrac{1}{\ell^{2}(4c_{1}\sqrt{\rho_{0}}r^{2}-\omega_{2})^{3}}[64c_{1}^{3}\rho_{0}^{\frac{3}{2}}r^{6}-48c_{1}^{2}\rho_{0}\omega_{2}r^{4}+\nonumber\\
    &12c_{1}\omega_{2}^{2}r^{2}\sqrt{\rho_{0}}-\omega_{2}^{3}+192c_{1}\ell^{2}r^{2}\rho_{0}^{\frac{3}{2}}+16\rho_{0}\omega_{2}\ell^{2}]\\
    \lambda_{2}=&-\lambda_{3}=\dfrac{16c_{1}^{2}\rho_{0}r^{4}-8c_{1}\omega_{2}\sqrt{\rho_{0}}r^{2}+\omega_{2}^{2}-16\rho_{0}\ell^{2}}{\ell^{2}(4c_{1}r^{2}\sqrt{\rho_{0}}-\omega_{2})^{2}}
\end{align}

For $r\to \infty$, then $\lambda_{1}=\lambda_{2}=-\lambda_{3}=1/\ell^{2}$. 
{The first extremum that is reached when approaching from infinity is located at $r_{rep}=\pm\sqrt{-\omega_{2}c_{1}\rho_{0}^{\frac{3}{2}}}/(2c_{1}\rho_{0})$
which corresponds to a local maximum of $\lambda_{2}$ and determines the place of the repulsion onset. On the other hand, $\lambda_{1}$ and $\lambda_{2}$ change sign at
\begin{align}
r^{(1)}_{dom}=&\pm\sqrt{\dfrac{\omega_{2}+4\ell \sqrt{\rho_{0}}}{4c_{1}\sqrt{\rho_{0}}}},\\
r^{(2)}_{dom}=&\dfrac{1}{2\sqrt{c_{1}}}\sqrt{\dfrac{\omega_{2}}{\sqrt{\rho_{0}}}+2(4B)^{\frac{1}{3}}-2\ell^2(\dfrac{16}{B})^{\frac{1}{3}}} ,
\end{align}
where
\begin{equation}
    B=\dfrac{\ell^2}{\sqrt{\rho_{0}}}\left(-\omega_{2}+\sqrt{\omega_{2}^{2}+4\rho_{0}\ell^2}\right) .
\end{equation}}
{According to our definition, we take the greatest value (the first zero from infinity) as indicating the region where repulsion dominates, i.e.,  $r_{dom}^{(2)}$. The weak  energy condition turns out to be identically satisfied for $r>0$. However, the strong energy condition is also violated at the radius $r_{0}$.}
{Notice the interesting relationship $r_{0} = r_{rep}/\sqrt{3}$, indicating that the repulsion radius can be associated with the location, where the strong energy condition is violated.} We conclude that the onset and  dominance region of repulsive gravity occur at radii close to the origin of coordinates. This result indicates that repulsion can be considered as inducing the regular behavior of the curvature at the center of the black hole.

\subsection{The second solution: $m=2,w_{1}=3$}

Again the underlying choice refers to the need of constant black hole mass. Specifically, for this case, the lapse function in Eq. \eqref{metric} is represented as follows
\begin{equation}\label{second}
	f(r)=c_{2}+\dfrac{r^2}{\ell^2}+\dfrac{2\rho_{0}}{\sqrt{c_{1}w_{2}\rho_{0}}}\arctan\left(\dfrac{2c_{1}\rho_{0}r^{2}}{\sqrt{c_{1}w_{2}\rho_{0}}}\right).
\end{equation}
The gravitational mass $M$ for a 2D sphere of radius $r$ becomes 
\begin{equation}
	M=2\pi\int_{0}^{\infty}r\rho(r)dr=-\dfrac{\pi^{2}\sqrt{-w_{2}\rho_{0}}}{\sqrt{c_{1}}w_{2}},
\end{equation}
which results in the following effective radius
\begin{equation}
R=\pm\sqrt{-\dfrac{\pi\sqrt{-w_{2}\rho_{0}}}{\sqrt{c_{1}}w_{2}}}.
\end{equation} 
For the effective radius above to make sense, i.e., $R>0$, we should demand $w_ 2<0$. The 
roots of the equation $f(r)=0$ determine the horizons of the above solution. In particular, 
the extremal root is obtained as
\begin{equation}
r_{ext}=\dfrac{\big(\rho_{0}^{3}c_{1}^{3}(w_{2}+4\rho_{0}\ell^2)\big)^{\frac{1}{4}}}{\sqrt{2}c_{1}\rho_{0}}.
\end{equation}
The left panel of Fig. \ref{fplot1} openly shows that $r_+>R$ for a large range of values of $c_1$, meaning that the solution in Eq.  \eqref{second} represents  a 3D black hole solution. Moreover, the curve $f(r)$ in the right panel also confirms that the second solution addresses a black hole (since $r_+>R$).

Regarding the asymptotic behavior of the metric \eqref{second} 
at $r\to\infty$ and $r\to 0$, we have
\begin{equation}\label{f}
	f(r)\sim c_{2}+\dfrac{\pi \sqrt{-\rho_{0}\omega_{2}}}{\sqrt{c_{1}}(-\omega_{2})^{\frac{3}{2}}}+\dfrac{r^2}{\ell^2}+\dfrac{1}{c_{1}r^{2}}+\mathcal{O}(r^{-6}),
\end{equation}
and
\begin{equation}\label{n}
	f(r)\sim c_{2}+\left(\dfrac{1}{\ell^2}+\dfrac{4\rho_{0}}{\omega_{2}}\right)r^{2}+\mathcal{O}(r^{6}),
\end{equation}
respectively.
Unlike the first solution, here $c_2$ singly does not play the role of mass, but the term $c_{2}+\frac{\pi \sqrt{-\rho_{0}\omega_{2}}}{\sqrt{c_{1}}(-\omega_{2})^{\frac{3}{2}}}=-\mathcal{M}$, which is effectively  associated with the ADT mass of the configuration.
The relation in Eq. \eqref{n} explicitly shows that, for the second solution, the effective cosmological constant is 
\begin{equation}
\Lambda_{eff}=-\dfrac{1}{\ell^2}-\dfrac{4\rho_{0}}{\omega_{2}},
\end{equation}
which is positive for  $\frac{4\rho_0 \ell^2}{\mid\omega_2\mid}>1$. In addition, the choice $c_{2}=1$ indicates the presence of dS core, which is a direct consequence of the power-law part of $P_t(\rho)$ in Eq. \eqref{eqq11}.

In the same manner, the Ricci and Kretschmann scalars  are both finite everywhere i.e., 
\begin{align}
	\lim_{r\to 0}\mathcal{R}=&-\dfrac{6}{\ell^2}-\dfrac{24\rho_{0}}{w_{2}}=6\Lambda_{eff},\\
	\lim_{r\to 0}\mathcal{K}=&\dfrac{12(w_{2}+4\rho_{0}\ell^{2})^{2}}{\ell^{4}w_{2}^{2}}=12\Lambda_{eff}^{2},
\end{align}

Similar to the first solution, from the right panel in Fig. \ref{fplot1}, one can conclude that the transverse WEC is met everywhere, while the SEC works for regions $r>r_0$
\begin{equation}
r_{0}=\pm\left({\dfrac{-w_{2}}{12c_{1}{\rho_{0}}}}\right)^{\frac{1}{4}},
\end{equation} 
namely, everywhere within the dS core except for the region near the origin with radius $r_0$. 

As before, Fig. \ref{Second} shows that around the origin the value of the anisotropic function vanishes, meaning that there is a vacuum in the center of the second solution.

\begin{figure*}[ht!]
	\centering
	\includegraphics[width=0.95\columnwidth]{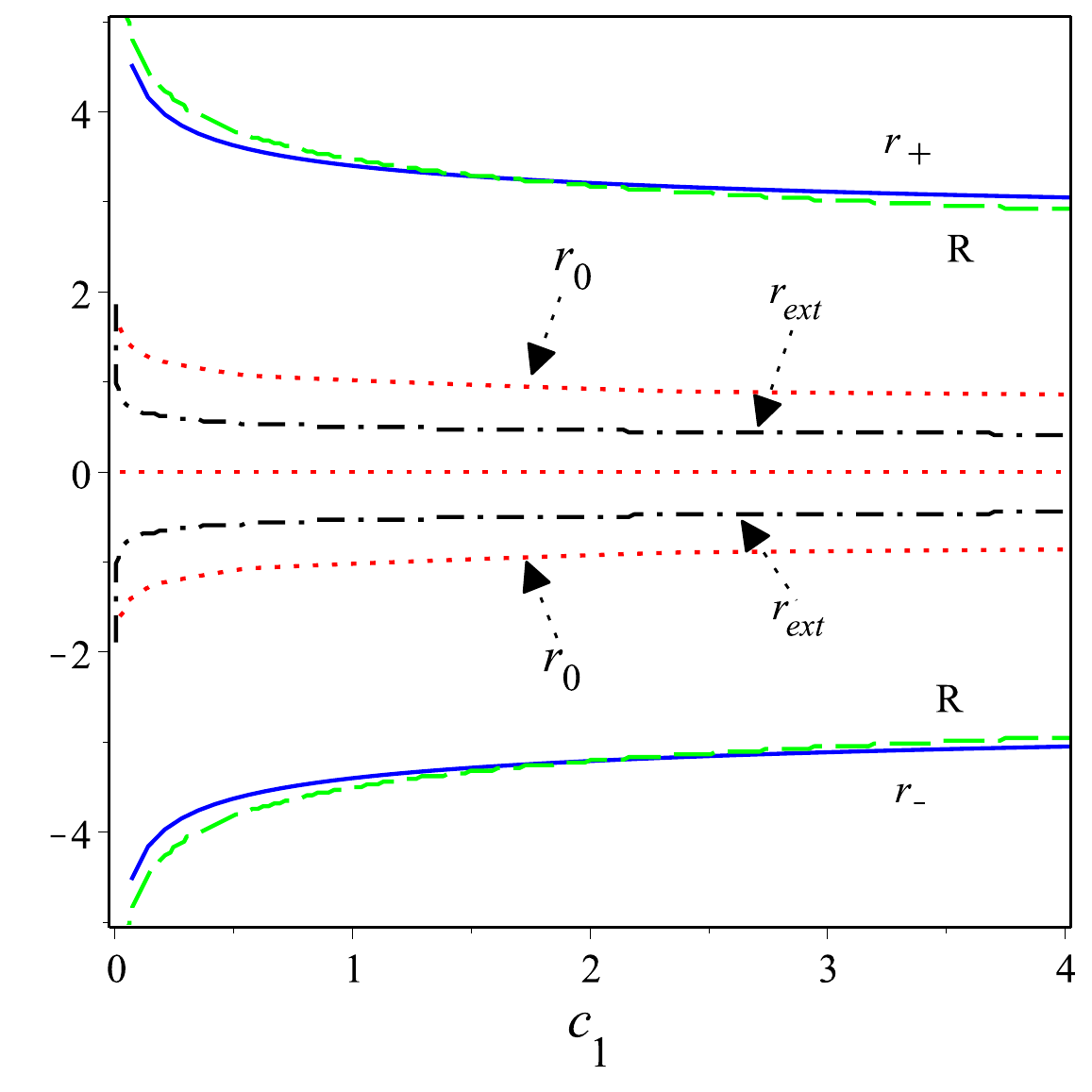}~~~~~~~~~~
	\includegraphics[width=0.95\columnwidth]{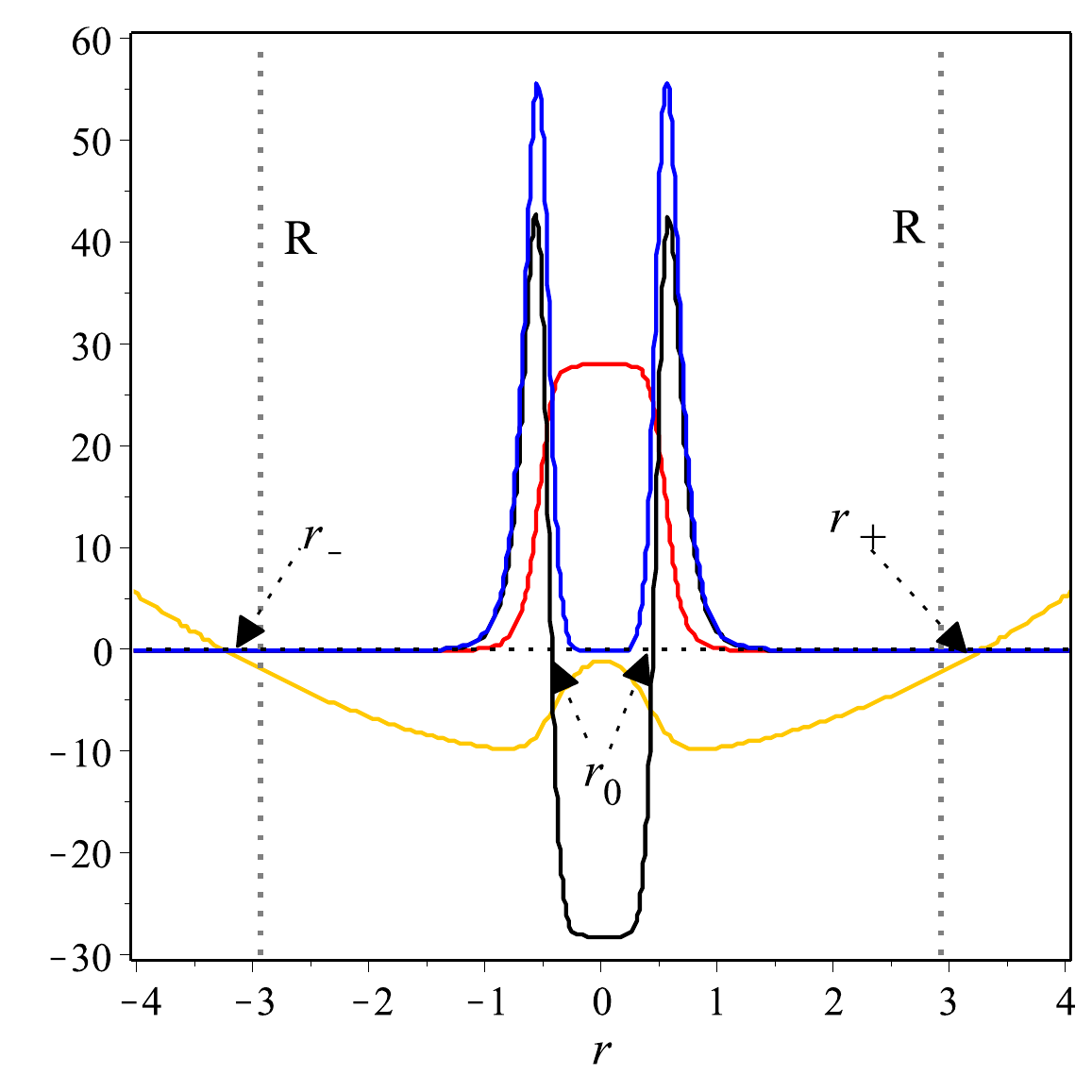}
	\caption{Same as Fig. \ref{rcplott} but for the third black hole solution with free parameters  $m=2, \omega_{1}=7, \omega_2=-1$, and $\rho_{0}=3.5$
	} 
	\label{wecL}
\end{figure*}

\begin{figure}[ht!]
	\centering
	\includegraphics[width=0.95\columnwidth]{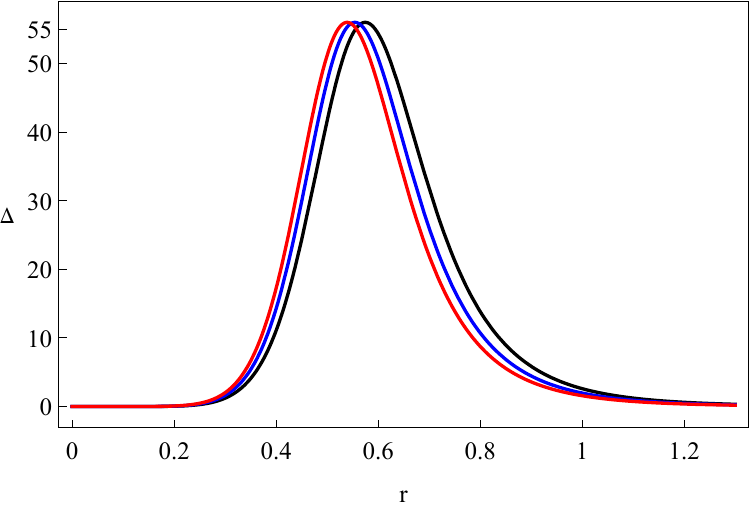}~~~
	\caption{The behavior of $\Delta-r$ for the third black hole solution with the same numerical values used in Fig. \ref{wecL}. 
	For the model parameter $c_1$, we have set numerical values $c_1=3,~3.5$, and $4$ in the black, blue, and red curves, respectively.} 
	\label{Third}
\end{figure}

{Computing the curvature eigenvalues $\lambda_{i}$ for this metric,  according to Appendix \ref{app1},  we can write}
\begin{align}
    \lambda_{1}&=\dfrac{\omega_{2}^{2}-8\omega_{2}c_{1}\rho_{0}r^{4}+16c_{1}^{2}\rho_{0}^{2}r^{8}+4\rho_{0}\omega_{2}\ell^2+48\rho_{0}^{2}c_{1}\ell^{2}r^{4}}{\ell^{2}(4c_{1}\rho_{0}r^{4}-\omega_{2})^{2}}\nonumber\\
    \lambda_{2}&=-\lambda_{3}=\dfrac{\rho_{0}(4c_{1}r^{4}-4\ell^{2})-\omega_{2}}{\ell^{2}(4c_{1}\rho_{0}r^{4}-\omega_{2})^{2}}
\end{align}

{The first extremum that is reached when approaching from infinity is located at $r_{rep}=\pm\frac{1}{c_{1}\rho_{0}}\left(\frac{-5\omega_{2}c_{1}^{3}\rho_{0}^{3}}{12}\right)^{\frac{1}{4}}$
which corresponds to a local maximum of $\lambda_{1}$. On the other hand, $\lambda_{1}$ and $\lambda_{2}$ change sign at
\begin{align}
  r^{(1)}_{dom}=&\pm\frac{\left(c_{1}^{3}\rho_{0}^{3}(\omega_{2}+4\rho_{0}\ell^{2})\right)^{\frac{1}{4}}}{\sqrt{2}c_{1}\rho_{0}},\\
  r^{(2)}_{dom}=&\dfrac{(-c_{1}^{3}\rho_{0}^{3}(-\omega_{2}+6\rho_{0}\ell^2-2\sqrt{9\rho_{0}^{2}\ell^4-4\rho_{0}\omega_{2}\ell^2}))^{\frac{1}{4}}}{\sqrt{2}c_{1}\rho_{0}} .
\end{align}
According to our definition, the greatest value indicates  the  region  where   repulsion  dominates, i.e. $r^{(1)}_{dom}$.  The WEC turns out to be identically satisfied for $r >0$. However, the SEC is violated at the radius $r_{0}$. As in the previous case, we find a relationship between the repulsion radius and the location where the SEC is violated, i.e, $r_{0}=r_{rep}/\sqrt{5}$. 
}

\subsection{The third solution: $m=2$, $w_{1}=7$}

Guaranteeing a suitable choice of constant that leads to a constant black hole mass, as above, we here find the lapse function to read  
\begin{align}\label{t3}
f(r)=&c_{2}+\dfrac{r^2}{\ell^2}+\dfrac{\rho_{0}A^{\frac{1}{4}}}{\sqrt{2}w_{2}}\ln\left[\dfrac{2r^{4}+\sqrt{2}r^2A^{\frac{1}{4}}+\dfrac{\sqrt{A}}{2}}{2r^{4}-\sqrt{2}r^2A^{\frac{1}{4}}+\dfrac{\sqrt{A}}{2}}\right]\\ \nonumber
&+\dfrac{\sqrt{2}\rho_{0}A^{\frac{1}{4}}}{w_{2}} \arctan\left(\dfrac{2\sqrt{2}r^2A^{\frac{1}{4}}}{A^{\frac{1}{2}}-4r^{4}}\right),
\end{align}
where $A=-2w_{2}/c_{1}\rho_{0}$. For a 2D sphere of radius $r$,
the gravitational mass $M$ and relevant effective radius $R$ can be expressed as follows
\begin{equation}
M=2\pi\int_{0}^{\infty}r\rho(r)dr=\dfrac{\pi^2 (2\rho_{0})^{\frac{3}{4}}}{c_{1}^{\frac{1}{4}}(-w_{2})^{\frac{3}{4}}},
\end{equation}
and
\begin{equation}
R=\pm\dfrac{\pi}{c_{1}^{\frac{1}{8}}}\left(\dfrac{2}{-w_{2}\rho_{0}}\right)^{\frac{3}{8}}.
\end{equation} 
As in the previous two cases, here we also need $c_1>0$ and $w_2<0$. Unlike the previous two cases, the lapse function in Eq. \eqref{t3} does not allow to present analytical expressions for $r_{\pm}$, and $r_{ext}$. Hence, we extract them in a numerical manner in Fig. \ref{wecL}, from which it is evident, and also from the behavior of $f(r)$ shown in the right panel, that  the third solution represents a  3D black hole solution.

The behavior of the lapse function \eqref{t3} as $r\to\infty$ and $r\to 0$ is
\begin{equation}\label{tf}
	f(r)\sim c_{2}+\dfrac{r^2}{\ell^2}+\dfrac{\sqrt{2}\pi\rho_{0}A^{\frac{1}{4}}}
	{w_{2}}+\mathcal{O}(r^{-6}),
\end{equation}
and
\begin{equation}\label{tn}
	f(r)\sim c_{2}+\left(\dfrac{1}{\ell^2}+\dfrac{8\rho_{0}}{w_{2}}\right)r^2+\mathcal{O}(r^{3}),
\end{equation} respectively. First, Eq. \eqref{tf} implies that the asymptotic limit of the third solution meets the lapse function of static BTZ black hole solution provided that $c_{2}+\frac{\sqrt{2}\pi\rho_{0}A^{\frac{1}{4}}}
{w_{2}}=-\mathcal{M}$. Given that the gravitational mass is dimensionless, so 
it can be easily recognized that $[c_1]=L^{-6}$, and $c_2$ is dimensionless.
Second, one can interpret the coefficient of $r^{2}$ in Eq. \eqref{tn} as an effective cosmological constant as
\begin{equation}
	\Lambda_{eff}=-\left(\dfrac{1}{\ell^2}+\dfrac{8\rho_{0}}{w_{2}}\right).
\end{equation}
As before, one can show that the Kretschmann and Ricci scalars are finite at the center of the black hole since
\begin{align}
&\lim_{r\to 0}\mathcal{R}=6\Lambda_{eff},\\
&\lim_{r\to 0}\mathcal{K}=\lim_{r\to 0}\mathcal{R}_{a b c d}\mathcal{R}^{a b c d}=12\Lambda_{eff}^{2}.
\end{align}
Here, as in the previous two cases, from Fig. \ref{wecL} (right panel) one can conclude that the WEC is satisfied everywhere, while the SEC is satisfied everywhere except near the origin at
\begin{equation}
r_{0}=\pm\Big(\frac{-c_{1}^3\rho_{0}^{3}w_{2}}{56}\Big)^{1/8}
(c_{1}\rho_{0})^{-1/2}.
\end{equation}
The equation above explicitly tells us that the GPEoS \eqref{eqq11} results in pushing the SEC physics into a small region deep in core, i.e., $r<r_{0}$, just like the previous ones. 

It is evident from Fig. \ref{Third} that the third black hole solution around the center $r=0$ 
 corresponds to a vacuum (i.e., $\Delta=0$), as in the previous two cases.

Now we analyze the behavior of repulsive gravity in this black hole. 
{The curvature eigenvalues, $\lambda_{i}$, are  given in terms of cumbersome expressions that cannot be represented in a compact form. However, a numerical study shows that the radii $r_{rep}$ and $r_{dom}$ are located in a region close to the center, as depicted in Fig. \ref{Third}.}

\begin{figure}[ht!]
	\centering
 \includegraphics[width=0.95\columnwidth]{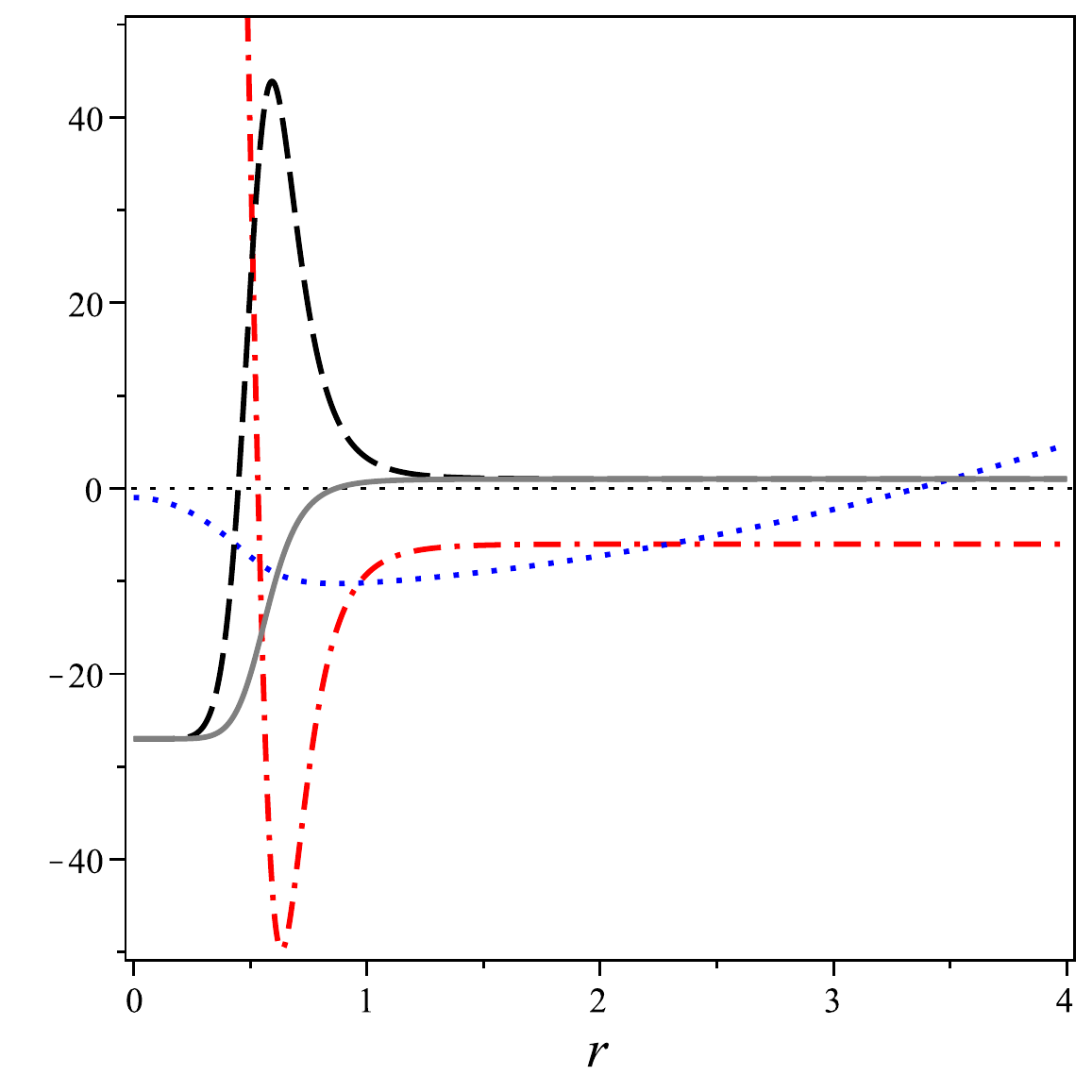}~~~
	\caption{Plots of the eigenvalues, $\lambda_{1}$,
$\lambda_{2}$ and $f(r)$, $R$ for the third solution for $c_{1}=3,\ell=1,\rho_{0}=3.5,\omega_{2}=-1$.} 
	\label{Third}
\end{figure}
We see that the eigenvalues are finite everywhere. The main result is that repulsive gravity is present in
both eigenvalues and, in fact, it becomes the dominant component as the origin of coordinates is approached, forcing the eigenvalues to become finite at $r = 0$. Thus, we conclude that the regular behavior of this black hole is a consequence of the presence of repulsive gravity.

Let us close this section by mentioning an important point.
Apart from the analytic solutions presented above, the existing setup allows large numbers of non-black hole solutions, which are obtained by setting other values for $m$ and $w_1$. However, we limit ourselves to pointing to one of them since the non-black hole solutions are beyond the scope of this work.  
For instance, setting the values $m=2,w_{1}=1$, we obtain a solution in which the total mass and, subsequently, the radius diverge from the view of an observer at infinity, meaning that this solution is unbounded and cannot represent a black hole.

\section{Geodesic completeness}
\begin{figure*}[ht!]
	\centering
	\includegraphics[width=0.95\columnwidth]{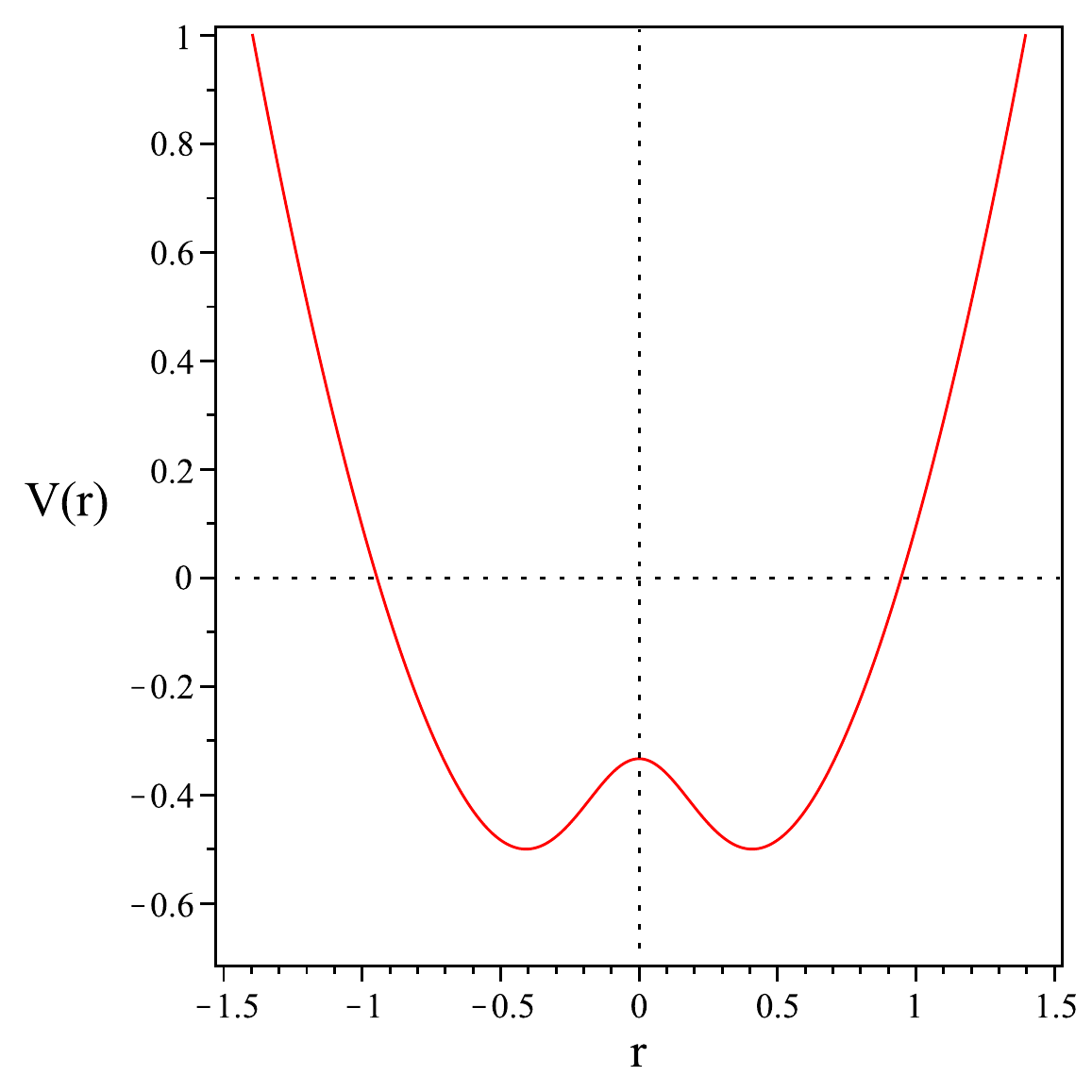}~~~~~~~~~~
	\includegraphics[width=0.95\columnwidth]{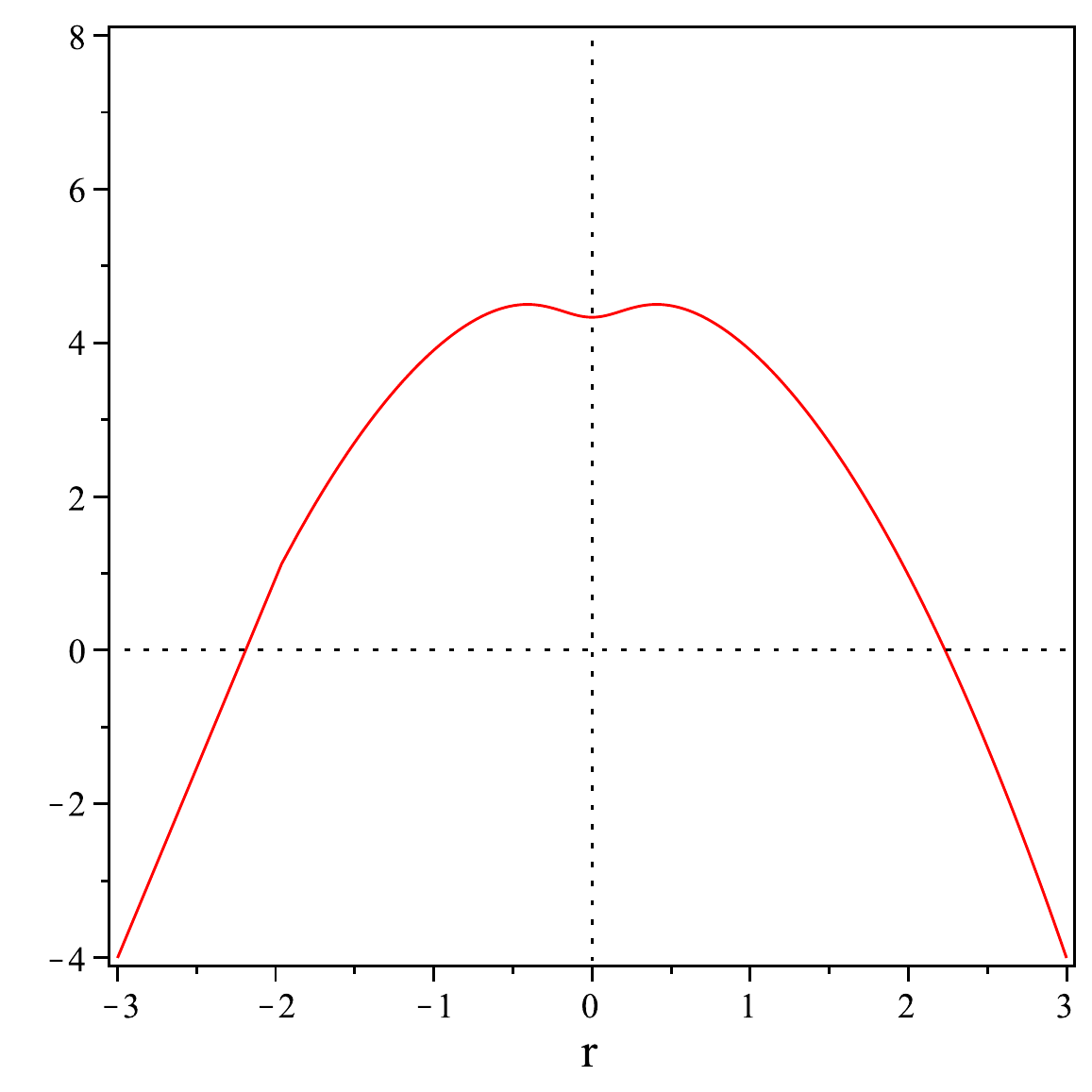}
	\caption{The left panel shows effective potential $V(r)$ for $m=2,\omega_{1}=3,\omega_{2}=-2,\rho_{0}=\ell=1$. The right panel shows $\dot{r}^2$ for the first solution with $m=2,\omega_{1}=3,\omega_{2}=-2,\rho_{0}=\ell=1,\epsilon=2$.
	} 
	\label{Vrplot1}
\end{figure*}

A powerful characterization of spacetimes containing singularities, is provided by the notion of geodesic completeness, namely, whether an affine parameter on every geodesic curve extends to arbitrarily large values or not. In a singular spacetime, there exist geodesic curves for which the affine parameter cannot be extended to arbitrarily large values, i.e., they start or terminate at a finite value of the affine parameter. 
Let us examine the photon and particle motions in the AdS spacetime. Take the static coordinates and consider a particle which starts from the bottom of the gravitational potential $r=0$ to $r=\infty$. The metric of space time is given by
\begin{equation}
ds^2=-f(r)dt^2+\dfrac{dr^2}{f(r)}+r^2d\phi^2.
\end{equation}
To find the geodesic equations, we take
\begin{equation}
    -f\dot{t}^2+\dfrac{\dot{r}^2}{f(r)}+r^2\dot{\phi}^2=k
\end{equation}
Using the constant of geodesics, namely $E=-f(r)\dot{t}$ and $L=r^2\dot{\phi}$ , we end up with the following equation for the geodesics,
\begin{equation}
  \dot{r}^2+V(r)=E^2,\;\;\;\;V(r)=f\left(\dfrac{L^2}{r^2}-k\right)  
\end{equation}
In the next section, we will study  stable circular orbits of null geodesics ($k=0$) under the constrains
\begin{equation}
    V\vert_{r_{p}}=V^{\prime}\vert_{r_{p}}=0,
\end{equation}
where $r_p$ is the radius of the photon sphere.
Here, to study the geodesic completion of spacetime we need to consider the radial geodesics, i.e. $L=0$. The radial geodesics satisfy the following equation in proper time,
\begin{equation}
\dot{r}^2=E^2+kf(r)
\end{equation}
First, consider the radial photon motion $k=0$, from the point of view of the proper affine parameter $\lambda$:
\begin{equation}
-\dfrac{E^2}{f(r)}+\dfrac{\dot{r}^2}{f(r)}=0\;\;\;\to\;\;\;\dfrac{dr}{d\lambda}=E
\end{equation} 
Thus, $r=E\lambda$, namely it takes an infinite affine parameter time until the photon reaches the AdS boundary. Therefore the space time for massless particles is geodesically complete.
Now, consider the particle motion from the point of view of the proper time $\tau$.
\begin{equation}\label{eq66rdot}
-\dfrac{E^2}{f(r)}+\dfrac{\dot{r}^2}{f(r)}=-1\;\;\;\to\;\;\;\dot{r}^2=E^2-f(r)
\end{equation}
For the  BTZ space time with $f(r)=c_{2}+\frac{r^2}{\ell^2}$, 
we have
\begin{equation}
\dot{r}^2=E^2-c_{2}-\frac{r^2}{\ell^2}
\end{equation}
For $E>\sqrt{c_{2}}$, the right-hand side of the above equation is positive at $r=0$. But the
right-hand side becomes negative as $r\to\infty$, so that the particle cannot reach the AdS
boundary $r\to\infty$. This is the effect of the gravitational potential well in the AdS
spacetime. The turning point $R$ is given by $R=\ell\sqrt{E^2-c_{2}}$. Then,
\begin{equation}\label{eq68tau}
\tau=\int_{0}^{R}\dfrac{dr}{\sqrt{E^2-c_{2}-\frac{r^2}{\ell^2}}}=\int_{0}^{\frac{\pi}{2}}d\varphi =\dfrac{\pi}{2}
\end{equation}
where $r=\ell\sqrt{E^2-c_{2}}\sin\varphi$. {Since the particle is reflected near the boundary and continues its path, the geodesic does not stop there. Therefore, the BTZ space time is geodesic complete for both massive and massles particles.}\\
Using \eqref{first} in \eqref{eq66rdot}, the radial geodesic equation of motion is given by
\begin{equation}
	\dot{r}^{2}=\epsilon^2 -c_{2}-\dfrac{r^2}{\ell^2}-\dfrac{4\sqrt{\rho_{0}}}{c_{1}(4c_{1}\sqrt{\rho_{0}}r^2-w_{2})}.
\end{equation}
while the effective potential reads:
\begin{equation}
	V(r)=c_{2}+\dfrac{r^2}{\ell^2}+\dfrac{4\sqrt{\rho_{0}}}{c_{1}(4c_{1}\sqrt{\rho_{0}}r^2-w_{2})}.
\end{equation}
The plots of $\dot{r}^2$ and $V$ are shown in Fig. \ref{Vrplot1}. As can be seen from the figures, for $c_{1}\omega_{2}<0$ ($\omega_{2}<0,c_{1}>0$) the metric and $\dot{r}^2$ are regular everywhere. 

Since the metric \eqref{first} is asymptotically AdS, the initial radius should be taken somewhere between the boundary of AdS and the black hole.
A probe particle with energy $\epsilon < c_{2}$ can not overcome the potential barrier and will bounce back consistently with the causality structure of the spacetime. On the the other hand, for $\epsilon > c_{2}$, $\dot{r}^2\to \epsilon^2-c_{2}+\frac{4\sqrt{\rho_{0}}}{c_{1}\omega_{2}}$ when $r\to 0$. Integrating equation \eqref{eq68tau} for the case of the metric \eqref{first}, it is easy to show that a probe particle can reach the point $r = 0$ in a finite amount of proper time. 

Moreover, the metric is symmetric under $r\to-r$, therefore the antipodal continuation is analytic \cite{Zhou:2022yio}. So, it is not necessary to extend the coordinate $r$ to negative values for such spacetimes. 
 As we have also shown that the affine parameter to approach $r = \infty$ diverges for massless particles. {Hence, the spacetime with (\ref{first}) is geodesics complete for massless and massive particles.}

{Similarly, the two other solutions are also geodesically complete for both massless and massive particles.}

\begin{table*}[ht!]
	\begin{center}
		{\hfill
			\hbox{	
		\begin{tabular}{|c|c|c|c|}
					\hline
					$c_1$&$\omega_2$&$r_{+}$&$r_{p}$\\										\hline
					$1$&$-1$&$0.208516$&$0.553223$  \\  \hline
					$2$&$-1$&$0.208518$&$0.517969$ \\  \hline
					$3$&$-1$&$0.20852$&$0.498539$ \\  \hline
					$4$&$-1$&$0.208522$&$0.485266$ \\  \hline
					$5$&$-1$&$0.208524$&$0.475257$ \\  \hline
					\end{tabular}
				}
			\hfill
		\hbox{
			\begin{tabular}{|c|c|c|c|}
			\hline
			$c_1$&$\omega_2$&$r_{+}$&$r_{p}$\\										\hline
			$2$&$-1$&$0.208518$&$0.517969$  \\  \hline
			$2$&$-2$&$0.301565$&$0.603835$ \\  \hline
			$2$&$-3$&$0.378253$&$0.66355$ \\  \hline
			$2$&$-4$&$0.448256$&$0.71179$ \\  \hline
			$2$&$-5$&$0.516065$&$0.753633$ \\  \hline
		\end{tabular}
	}
\hfill
}
			\caption{Numerical values of $r_+$, and $r_p$ for the third solution \eqref{t3} in which fixed the different values for $c_1$, and $\omega_2$. In the left table, the values of $c_1$ change while the value of $\omega_2$ remains unchanged. In the right table, the opposite occurs. 
			We also use $\ell=1$, and $\rho_0=3.5$. }
			\label{table}
			\end{center}
		\end{table*}

\section{On the shadows of the solutions} \label{shadow}
In this section, we take a step to understand the kinematics of photons around these 3D black hole solutions at hand. 
We do it by quantifying the shadow that appears as photons move in the gravitational field described by these  solutions. In this regard, the photon sphere is an essential concept closely related to the black hole shadow \cite{Vagnozzi:2022moj}. A black hole can potentially cast a shadow provided that a geometrically thick and optically thin emission region forms outside the event horizon  \cite{Falcke:1999pj}. In other words, to survey the optical observations such as black hole shadows via background light emission, the formation of a 
photon sphere around the black hole is essential \footnote{This may be true for black hole spacetimes that have an event horizon. In contrast, the existence of a photon sphere is not necessary at all for the shadow to appear around other compact objects without an event horizon, such as  naked singularities \cite{Joshi:2020tlq}. It means that compact objects other than black holes are also prone to produce  shadows.}.

In similarity of 4D, to study the null geodesics around 3D black hole solutions, one can utilize the standard Hamilton-Jacobi method \cite{chand}. By skipping details, for a 3D static, spherically symmetry spacetime, the radius of the shadow is obtained as follows (for more details see, for example \cite{Upadhyay:2023yhk})
\begin{equation}\label{1116}
R_s=\frac{r_{p}}{\sqrt{f(r_{p})}},
\end{equation} 
where $r_{p}$, in essence, is the radius of photon sphere
and it is given by solving the following equation
\begin{equation}\label{1117}
r_{p}f^{\prime}(r_{p})- 2f(r_{p})=0.
\end{equation}
By taking into account the first and second solutions (i.e., lapse functions \eqref{first}, and \eqref{second} ) in Eq. \eqref{1117}, we find that 
Eq. \eqref{1117} does not have real solutions for any of the possible values of the involved parameters. This means that the 3D black hole solutions \eqref{first} and \eqref{second} do not allow the presence of a photon sphere, implying that photons around these 3D black holes  cannot form circular orbits and are indeed scattered.
As a result, these two solutions do not cast shadows.

This is not the case for the third solution (i.e., the lapse function \eqref{t3}) because it allows the presence of a  photon sphere for some values of the involved parameters. In Table \ref{table}, we present some numerical solutions, which show  clearly that the location of the photon sphere of the third solution is outside the event horizon, as expected. However, by inserting these numerical solutions in \eqref{1116}, we see that no real solutions for $R_s$ can be obtained \footnote{Note that with other combinations of $(c_ 1$, $\omega_ 2)$, except for those given in Table \ref{table}, we can generate many allowed values for $r_{p}$, i.e., with $r_ p>r_+$. However, this will not change the final result regarding the absence of shadows.}. It means that despite the presence of the photon sphere in the 3D black hole solution \eqref{t3}, it will not lead to the formation of a shadow.
In general, it can be said that none of these 3D black hole solutions at hand cast a shadow. If 3D shadows would form, one could via encountering with ''Event Horizon Telescope'' data,  extract some constraints for the free parameters of these solutions.

\section{Mass inflation}\label{eqsec5}

The metric \eqref{metric} in ingoing Eddington-Finkelstein coordinate $v$ using $dt=dv+dr/f$ is given by the line element
\begin{equation}\label{eqmi73}
    ds^2=-f(v,r)dv^2+2dvdr+r^2d\phi^2
\end{equation}
This metric describes a solution where a field is  collapsing at the speed of light along null curves. Before considering the multipolytropic Vaidya solutions, we assume that all components of $T_{\mu\nu}$ vanish except for
\begin{equation}\label{eqmi74}
    T_{vv}=\dfrac{\epsilon(v)}{2\pi r}.
\end{equation}
Inserting the metric (\ref{eqmi73}) with assumption (\ref{eqmi74}) in the field equation, we have 
\begin{align}
G_{\phi\phi}+\Lambda g_{\phi\phi}=0,\;\;\;\to\;\;\;\dfrac{\partial^{2}f(v,r)}{\partial r^{2}}+2\Lambda =0
\end{align} 
solving the above give us
\begin{equation}\label{eqmi3}
f(v,r)=F_{1}(v)+F_{2}(v)r-\Lambda r^2
\end{equation}
we assume $F_{2}(v)=0$. Inserting \eqref{eqmi3} in the $vv$-component of the field equation and assuming $\epsilon(v)=\mu\delta(v-v_{0})$ we have
\begin{equation}
G_{vv}+\Lambda g_{vv}=8\pi T_{vv},\;\;\;\to\;\; \dfrac{dF_{1}(v)}{dv}-8\mu\delta(v-v_{0})=0
\end{equation}
solving it for $F_{1}(v)$, we have
\begin{equation}
F_{1}(v)=c_{1}+8\mu\theta(v-v_{0})
\end{equation}
Finally we have 
\begin{equation}
f(v,r)=c_{1}+8\mu\theta(v-v_{0})-\Lambda r^2
\end{equation}
This solution is the Vaidya shockwave in 2+1 dimensions. So,
\begin{itemize}
\item for $v<v_{0}$, we have an AdS$_{3}$ space with metric
\begin{equation}
ds^2=-(c_{1}-\Lambda r^2)dv^2+2dvdr+r^2d\phi^2,
\end{equation}
\item for $v>v_{0}$, we have the BTZ metric
\begin{equation}
ds^2=-(c_{1}+8\mu-\Lambda r^2)dv^2+2dvdr+r^2d\phi^2.
\end{equation}
\end{itemize}
Therefore, by glueing the Penrose diagrams for the AdS$_{3}$ and  BTZ spacetimes, 
the Vaidya shockwave geometry is obtained \cite{Han:2024unm}.


Now, we consider the multipolytropic Vaidya metric of solution \eqref{first},
\begin{equation}\label{mifirst}
	f(v,r)=c_{2}(v)+\dfrac{r^2}{\ell^2}+\dfrac{4\sqrt{\rho_{0}}}{c_{1}(4c_{1}\sqrt{\rho_{0}}r^2-w_{2})}.
\end{equation}
Here, the arbitrary mass function $c_{2}(v)$ determines the flux of incoming radiation.
As  mentioned before in Eq. n \eqref{eqf22}, this solution has two horizons under the condition:
\begin{equation}
    Y>0,\;\;\;\;\;\;c_{2}<0.
\end{equation}
To study the stability of the Cauchy horizon, we follow the Ori approach \cite{Ori:1991zz}, in which the outgoing energy flux is modelled by a thin pressureless null shell $\Sigma$ which divides the spacetime into two regions inside and outside the shell. The metric in each sector of spacetime can be written as \cite{Carballo-Rubio:2022kad,Bonanno:2020fgp}
\begin{equation}
    ds^2=-f_{\pm}(r,v_{\pm})dv_{\pm}^2+2drdv_{\pm}+r^2d\phi^2
\end{equation}
Since the shell moves light-like, we have 
\begin{equation}
    f_{-}dv_{-}=2dr.
\end{equation}
Continuity of the flux across the shell requires
\begin{equation}
[T_{\mu\nu}s^{\mu}s^{\nu}]=0.    
\end{equation}
In terms of metric function, it is easy to show that
\begin{equation}
    \dfrac{1}{f_{+}^{2}}\dfrac{df_{+}}{dv_{+}}=\dfrac{1}{f_{-}^{2}}\dfrac{df_{-}}{dv_{-}}
\end{equation}
here we have used $s^{\mu}=(2/f_{\pm},1,0)$. The above equality can be re-written as follows
\begin{equation}
    \dfrac{1}{f_{+}}\dfrac{df_{+}}{dv_{+}}=F(v)
\end{equation}
where
\begin{equation}\label{eqq88q}
    F(v)=\dfrac{1}{f_{-}}\dfrac{df_{-}}{dv_{-}}
\end{equation}
here, we have used $f_{+}dv_{+}=f_{-}dv_{-}$. Defining the position of the shell by $R(v)$, we can get
\begin{equation}\label{eqmi89}
    \dfrac{dR}{dv}=\dfrac{1}{2}f_{-}
\end{equation}
Therefore, the metric function in each sector is given by
\begin{equation}
    f_{\pm}(v_{\pm})=c_{2\pm}(v_{\pm})+\dfrac{r^2}{\ell^2}+\dfrac{4\sqrt{\rho_{0}}}{c_{1}(4c_{1}\sqrt{\rho_{0}}r^2-w_{2})}.
\end{equation}
By substituting $f_{-}$ in \eqref{eqmi89}, the position of the shell is obtained
\begin{equation}
    \dfrac{dR}{dv}=\dfrac{1}{2}c_{2-}(v_{-})+\dfrac{r^2}{2\ell^2}+\dfrac{2\sqrt{\rho_{0}}}{c_{1}(4c_{1}\sqrt{\rho_{0}}r^2-w_{2})}
\end{equation}
For the mass function $c_{2+}$, we have
\begin{equation}\label{eqmi92}
    F(v)=\dfrac{\ell^2 c_{1}(\omega^2-4c_{1}\sqrt{\rho_{0}}R(v))\dfrac{dc_{2+}}{dv_{+}}}{c_{2+}\ell^2c_{1}\omega_{2}+c_{1}\omega_{2}R^2-4\sqrt{\rho_{0}}(\ell^2+c_{1}^2R^4+\ell^2 c_{1}^{2}R^2c_{2+})}
\end{equation}
with 
\begin{equation}\label{eqmi93}
 \dfrac{\ell^2 c_{1}(\omega^2-4c_{1}\sqrt{\rho_{0}}R(v))\dfrac{dc_{2-}}{dv_{-}}}{c_{2-}\ell^2c_{1}\omega_{2}+c_{1}\omega_{2}R^2-4\sqrt{\rho_{0}}(\ell^2+c_{1}^2R^4+\ell^2 c_{1}^{2}R^2c_{2-})}=F(v)   
\end{equation}
Now, we assume the following boundary condition for the mass function 
\begin{equation}\label{eqmi94}
    c_{2-}=c_{2}-\dfrac{\beta}{(\frac{v}{v_{0}})^p}
\end{equation}
and considering the asymptotic behavior of the shell as follows
\begin{equation}
    R(v)=r_{-}+\dfrac{1}{v^{s}}\sum_{i} \dfrac{a_{i}}{v^i},
\end{equation}
the coefficients of $a_{i}$ can be obtained as follows
\begin{equation}\label{eqmi96}
    R(v)\sim r_{-}-\dfrac{\beta}{2\kappa_{-}v^{p}}+\cdots,
\end{equation}
where $\kappa_{-}$ is the positive surface gravity of the Cauchy horizon ($\kappa_{-}=-f^{\prime}(r_{-})/2$).
By substituting \eqref{eqmi96} and \eqref{eqmi94} in \eqref{eqmi93}, we get\\
\begin{equation}\label{eqmi97}
   F(v)\sim \zeta pv^{p}+\cdots
\end{equation}
Finally, by substituting \eqref{eqmi97} and \eqref{eqmi96} in \eqref{eqmi92}, we obtain a finite expression for $c_{2+}$ in the limit $v\to \infty$ as follows:
\begin{equation}
    c_{2+}={c}_{2}+\mathfrak{c}_{2}e^{-\zeta v^{p}}.
\end{equation}
Here $\mathfrak{c}_{2}$ is an integration constant and $\zeta$ is a positive constant given by
\begin{equation}
    \zeta=\dfrac{16\ell^2\sqrt{\rho_{0}}\kappa_{-}^2}{\beta c_{1}(c_{2}\ell^2+r_{-}^2)\eta},
\end{equation}
here
$\eta=4\sqrt{\rho_{0}}c_{2}c_{1}\ell^2+16\sqrt{\rho_{0}}\ell^2 c_{1}r_{-}\kappa_{-}-\omega_{2}+24c_{1}r_{-}^2\sqrt{\rho_{0}}$.
Therefore, one can conclude that the mass of Cauchy horizon does not grow up and the first solution does not experience mass inflation. 

\begin{figure}[ht!]
	\centering
 \includegraphics[width=0.95\columnwidth]{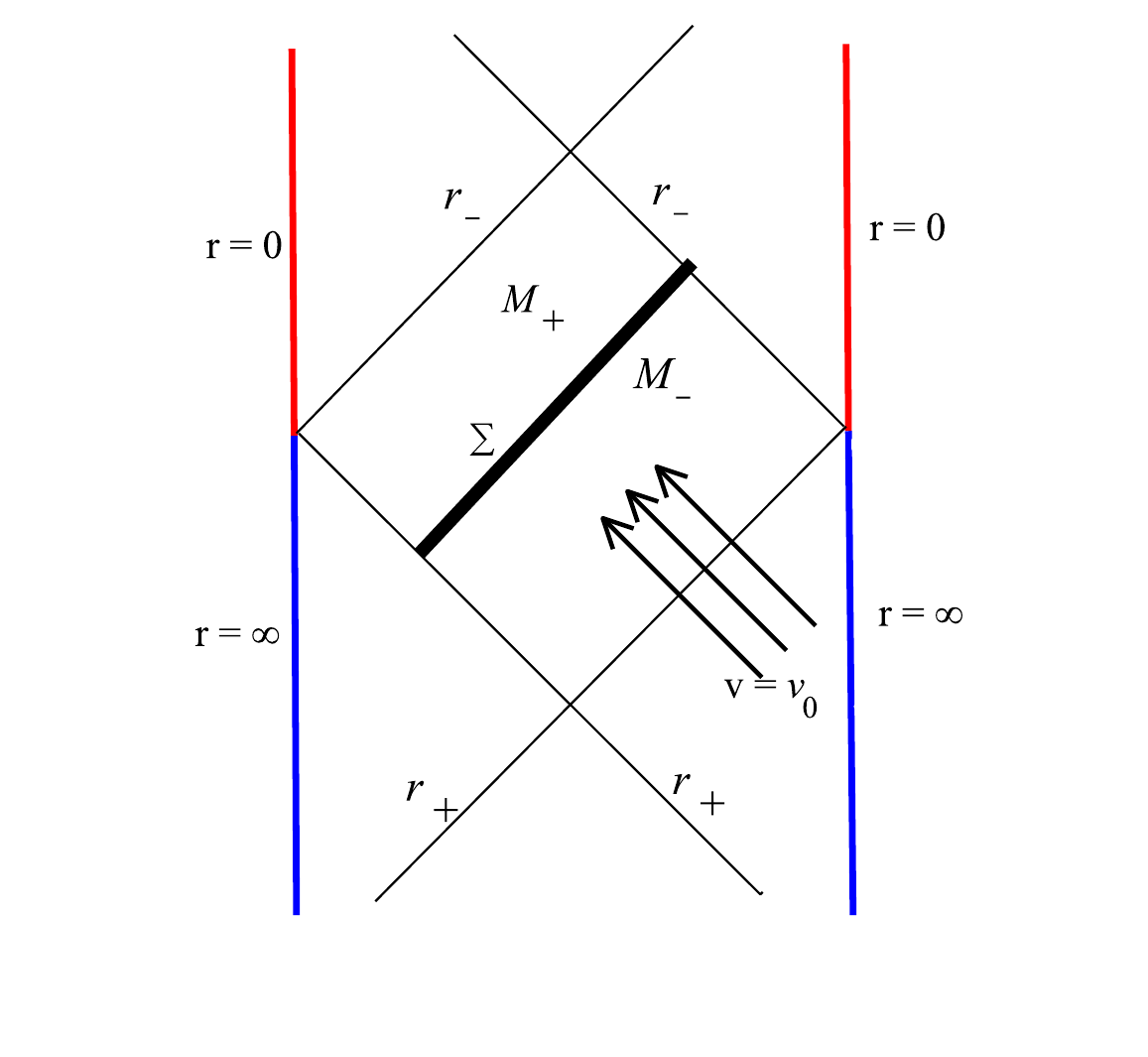}~~~
	\caption{The Penrose diagram of a regular-AdS black hole. The radiation entering the black hole is backscattered from the outer potential barrier and appears as incoming null rays. The thin shell $\Sigma$ separates spacetime into an inner region $M_{+}$ and an outer region $M_{-}$.} 
	\label{Third}
\end{figure}

For the other two solutions using the same method, it is straightforward to show that their Cauchy horizons are stable under the energy flow.


\section{Conclusion}\label{concluding}
Considering an anisotropic relativistic configuration of polytropes, described by the GPEoS, Eq. \eqref{eqq11}, in a 3D static background, we first generalized the TOV equations to derive compatible solutions. 

We found out that the structure of the anisotropic pressure consists of radial and transverse components, exhibiting vacuum energy and a GPEoS with a linear term plus a power-law contribution. This scenario is widely adopted in astrophysical contexts since it is thought to be well-approximate real compact objects. 

Consequently, we solved the full system of existing equations by setting certain values for the model parameters denoted as $\ell$, $c_1$, and $c_2$. Thus, we derived three analytical new black hole solutions prompting a dS core, i.e., being regular at $r = 0$ and smoothly approaching an asymptotically AdS spacetime. As mentioned before, the important point is that the origin of the dS vacuum inside the black hole comes from the power-law part of 
$P_{t}(\rho)$ in GPEoS, Eq. (\ref{eqq11}), i.e. $\omega_{2}$. In other words, the regularity of the black hole is due to the polytropic term.

The status of the WEC and SEC was discussed for the three solutions, by varying the above parameters. We underlined that both WEC and SEC are satisfied everywhere with the remarkable exception of the SEC that is violated in a small region inside the dS core. Physically, the existence of a dS core can be interpreted as the absence of singularity, suggesting the black hole to be regular.
We checked the regularity of the black hole solutions using the finiteness of curvature scalars and their geodesic completeness. We also showed that the Cauchy horizon of the solutions is stable using the mass inflation formalism.


Furthermore, by analyzing the behavior of the curvature eigenvalues, we demonstrated that repulsive gravity is present near the center of the black holes and it can interpreted as being responsible or the avoidance of curvature singularities. Thus, we can also conclude that the regularity property of the analyzed black holes is due to the presence of repulsive gravity of dS core. In the end, we conclude that regular black holes can be described by polytropic models similar to other compact objects due to the absence of singularities. Of course, with the difference that these solutions at hand do not cast shadows.

Finally, the solutions we are presenting are generalization of BTZ BH with a new matter component and the same topology. Therefore, since BTZ has a conical singularity, any generalization with only a matter field will have the same conical singularity. But this is not at the level of the curvature, but at the level of the metric (or/and the connection). The conical singularity can be removed by considering quantum fields and quantum perturbations of BH, which we are not considering in the present work.

As a perspective, we notice that these solutions can be further investigated, for instance, regarding the dynamic stability of these solutions against superradiance scattering and thermodynamics of the black hole solutions.

\section* {Acknowledgements}
SNS would like to thank F. Saueressig, N. Riazi and V. Taghiloo for insightful discussions. The work of OL is  partially financed by the Ministry of Education and Science of the Republic of Kazakhstan, Grant: IRN AP19680128.

\appendix

\section{{Repulsive gravity}}\label{app1}

In this Appendix, we review an invariant formalism for describing repulsive gravity in terms of the eigenvalues of the Riemman curvature tensor. 

To obtain the Riemann curvature tensor in a local orthonormal frame, we choose the orthonormal tetrad $e^a$ as
\begin{equation}
	ds^2=\eta_{a b}e^{a}\otimes e^{b}
\end{equation}
with $\eta_{ab}=diag(-1,1,1)$. The first 
\begin{equation}\label{eqq2}
	de^{a}+\omega^{a}{}_{b}\wedge e^{b}=0
\end{equation}
and second Cartan structure equations  
\begin{equation}\label{eqq3}
	\Omega^{a}{}_{b}=d\omega^{a}{}_{b}+\omega^{a}{}_{c}\wedge\omega^{c}{}_{b}=\dfrac{1}{2}R^{a}{}_{b c d}e^{c}\wedge e^{d},
\end{equation}
allow us to introduce the connection 1-form $\omega^a_{\ b}$ and the curvature 2-form  
$\Omega^a_{\ b}$, which determines the tetrad components of the curvature tensor $R^a_{\ bcd}$.

Furthermore, to compute the curvature eigenvalues, it is convenient to introduce the 
convention
\begin{equation}
	1\to 12,\;\;\;\;2\to 13,\;\;\;3\to 23.
	\label{conv}
\end{equation}
Then, the curvature eigenvalues are easily computed as the eigenvalues of the matrix $R_{AB} $.

Consider now the case of a
spacetime geometry in circularly symmetric 3D with  coordinates ($t, r, \phi$), 
\begin{equation}
	ds^2=-f(r)dt^2+\dfrac{dr^2}{f(r)}+r^{2}d\phi^2\;,
\end{equation}
where $t,r\in R$, $\phi \in [0,2\pi]$.
Therefore, we have
\begin{equation}
	e^{1}=\sqrt{f}dt,\;\;\;e^{2}=\dfrac{dr}{\sqrt{f}},\;\;\;e^{3}=rd\phi
\end{equation}
The components of the connection 1-form are  determined using \eqref{eqq2}, as follows
\begin{equation}
	\omega^{1}{}_{2}=\dfrac{f^{\prime}}{2\sqrt{f}}e^{1},\;\;\;\omega^{2}{}_{3}=-\dfrac{\sqrt{f}}{r}e^{3} \ .
\end{equation}
Moreover, using equation \eqref{eqq3}, we obtain
\begin{equation}
	R_{1212}=\dfrac{f^{\prime\prime}}{2},\;\;\;\;R_{1313}=\dfrac{f^{\prime}}{2r},\;\;\;R_{2323}=-\dfrac{f^{\prime}}{2r}.
\end{equation}
Finally, according to the convention (\ref{conv}), 
the ($3\times 3$)-matrix $R_{AB}$  can be written as
\begin{center}
	\begin{equation*}
		R_{AB}=  \begin{bmatrix}
			\dfrac{f^{\prime\prime}}{2} & 0&0 \\
			0 & \dfrac{f^{\prime}}{2r}&0\\
			0 & 0&-\dfrac{f^{\prime}}{2r}
		\end{bmatrix}
	\end{equation*}
\end{center}
Therefore, the curvature eigenvalues, $\lambda_{i}$, are given by \cite{Luongo:2023xaw,Luongo:2023aib,Luongo:2014qoa},
\begin{equation}
	\lambda_{1}=\dfrac{f^{\prime\prime}}{2},\;\;\;\lambda_{2}=\dfrac{f^{\prime}}{2r},\;\;\;\lambda_{3}=-\dfrac{f^{\prime}}{2r}.
\end{equation}
These eigenvalues encapsulate all the information regarding curvature and exhibit scalar behavior under coordinate transformations. The identification of repulsive gravity is established as follows: \emph{The change in sign of at least one eigenvalue signifies the transition to repulsive gravity, meaning the zeros of the eigenvalues mark the locations where repulsive gravity becomes predominant}. 

The existence of an extremum in an eigenvalue indicates the initiation of repulsive gravity, with the onset determined by the condition 

\begin{equation}
	\partial \lambda_{i}/\partial r=0.    
\end{equation}

\section{Mass Inflation of Charged-Static BTZ}
The metric functions of charged-static BTZ black hole is given by
\begin{equation}
    f(r)=\dfrac{r^2}{\ell^2}-M-\dfrac{Q^2}{2}\ln\left(\dfrac{r}{r_{0}}\right)
\end{equation}
using the approach of section \eqref{eqsec5}, and assumption \eqref{eqmi94}, it is easy to obtain the asymptotic behavior of the shell 
\begin{equation}
    R(v)=r_{-}+\dfrac{\beta}{2\kappa_{-}v^{p}}+\cdots
\end{equation}
substituting above equation into the equation \eqref{eqq88q} yields the asymptotic behavior of $F(v)$ as 
\begin{equation}
    F(v) =-\dfrac{8pr_{-}^2\kappa_{-}^2v^{p-1}}{\beta(4r_{-}^2+Q^2\ell^2)}+\cdots
\end{equation}
The asymptotic behavior of $M_{+}$ can then be obtained as follows
\begin{equation}
    M_{+}(v)=M+c\;e^{-\eta v^{p}}
\end{equation}
here $\eta$ is a positive constant as follows
\begin{equation}
   \eta=\dfrac{16r_{-}^2\ell^2\kappa_{-}^2}{\beta (Q^2\ell^2+4r_{-}^2)}
\end{equation}
Therefore, as $v\to \infty$, the $M_{+}$ approaches a constant value. 
 

\end{document}